\documentclass[preprints,article,submit,pdftex,moreauthors]{Definitions/mdpi} 

\firstpage{1} 
\pubvolume{1}
\issuenum{1}
\articlenumber{0}
\pubyear{2023}
\copyrightyear{2023}
\datereceived{ } 
\daterevised{ } 
\dateaccepted{ } 
\datepublished{ } 
\hreflink{https://doi.org/} 

\usepackage{multicol}
\preto{\abstractkeywords}{\nolinenumbers}

\Title{A symbolic approach to discrete structural optimization using quantum annealing}

\TitleCitation{Symbolic Structural Optimization with Quantum Annealing}

\Author{Kevin Wils and Boyang Chen$^{1}$*}

\AuthorNames{Kevin Wils and Boyang Chen}

\AuthorCitation{Wils, K.; Chen, B.}

\address{%
$^{1}$ \quad Department of Aerospace Structures and Materials, Faculty of Aerospace Engineering, Delft University of Technology. Kluyverweg 1, 2629 HS, Delft, The Netherlands}

\corres{Correspondence: B.Chen-2@tudelft.nl}

\abstract{With the advent of novel quantum computing technologies, and the knowledge that such technology might be used to fundamentally change computing applications, a prime opportunity has presented itself to investigate the practical application quantum computing. The goal of this research is to consider one of the most basic forms of mechanical structure, namely a 2D system of truss elements, and find a method by which such a structure can be optimized using quantum annealing. The optimization will entail a discrete truss sizing problem - to select the best size for each truss member so as to minimize a stress-based objective function. To make this problem compatible with quantum annealing devices, it will be written in a QUBO format. This work is focused on exploring the feasibility of making this translation, and investigating the practicality of using a quantum annealer for structural optimization problems. Using the methods described, it is found that it is possible to translate this traditional engineering problem to a QUBO form and have it solved by a quantum annealer. However, scaling the method to larger truss systems faces some challenges that would require further research to address. }


\begin{document}




\section{Introduction}

Quantum computers are rather unique devices that, by leveraging quantum mechanical principles, theoretically allow certain types of problems to be solved much more efficiently than is possible with classical computers \cite{nielsen_chuang_2010}. While classical computers use binary bits, 1s and 0s, to perform their computations, quantum computers make use of \textit{quantum bits}. Quantum bits, or \textit{qubits}, can not only represent the classical 0 and 1 states, but can also exist in a quantum superposition of these states. This quantum superposition, when leveraged effectively, is one of the reasons why quantum computers promise better performance in certain applications. 

There are two main types of quantum computers currently in development, being the General Purpose Quantum Computer (GPQC) and the Quantum Annealer (QA). With the GPQC, most of the potential improvements stem from the fact that these systems can run complex quantum algorithms, allowing for more efficient problem-solving methods to be devised. An overview of quantum algorithms is given by Montanaro \cite{Montanaro2016}. On the other hand, a QA can only use the quantum annealing algorithm to solve very specific types of optimization problems, known as quadratic unconstrained binary optimization (QUBO) or Ising model problems \cite{Dwave_gettingstarted}. The QA is said to leverage \textit{quantum tunneling} to aid in finding optimal solutions to optimization problems \cite{Denchev2016}. 

Both quantum computing technologies are still quite novel, and quantum computing hardware is still in its infancy compared to the advanced state of classical computing technologies. Research into practical applications of quantum computing stem only from the last several years, however, some studies have already shown promising results \cite{Zheng2017,Neukart2017,Vreumingen2019,Stollenwerk2019,van2022hybrid,ye2023quantum,lee2023size,sandt2023quantum}. A review of applications on structural mechanics is given by Tosti Balducci et al. \cite{tosti2022review}. Furthermore, in industry, some companies are already pushing for the development of early practical applications of quantum computing \cite{bayerstadler2021industry,bova2021commercial}. For example, Airbus has posted the Airbus Quantum Computing Challenge. One of the challenges is to optimize a wingbox structure, the main load-bearing component in aircraft wings, using a quantum computer \cite{Airbus2019}. Another example comes from Volkswagen, who have researched how a QA can be used to optimize a traffic flow problem \cite{Neukart2017}. Volkswagen has already applied this research for the real-time optimization of public transport routes in Lisbon \cite{Volkwagen2019Lisbon}. 

In the aerospace industry there is a continuous demand to develop the most lightweight structures, as this can lead to increased fuel efficiency, or increased payload capacity, both of which can lead to higher profits. This research will investigate the concept of using a quantum computer to assist in the optimization of mechanical structures. More specifically, the optimization of 2D truss structures will be targeted, as they are among the simplest structures to start with and are easily scalable to arbitrarily complex forms. Of the two types of quantum computers, the QA is chosen due to the higher level of technological maturity compared to GPQC, offering significantly more qubits of processing power. The main commercial supplier of QA technology is the company D-Wave Systems Inc. In this research, the D-Wave 2000Q quantum annealer, which has roughly 2000 qubits available, is used. During the execution of this research, a monthly allowance of 60 seconds of quantum processing unit (QPU) access time was available. This limit was taken into account for the scope of the testing that was performed. Because the QA can only solve specific types of optimization problems, the objective in this work will be to investigate how the truss optimization problem can be made compatible with the QA.

The rest of the paper will start with a brief description of the background on QA in Section~\ref{sec:backgroundQA}.  


\section{Background on QA: QUBO and Ising Formulations}
\label{sec:backgroundQA}
For any problem that one wishes to solve using a QA, the first step is to ensure the problem is written with either a QUBO or an Ising formulation. In both cases, the QA attempts to find a solution for which the Hamiltonian energy is minimal. 

To define a QUBO problem, an $N\times N$~matrix~$\textbf{Q}$ is needed. The matrix $\textbf{Q}$ is typically written as an upper triangular matrix. The QA attempts to find the optimal binary bitstring $\textbf{x}$ of length $N$ that minimizes the Hamiltonian energy $H$, as shown in \cref{eq:QUBO} \cite{Glover2018}. When a solution to the QUBO problem is found, the binary variables in the solution vector $\textbf{x}$ will then have values of either 0 or 1.   

\begin{equation}
\begin{aligned}
\min{\left(H\right)} &= \textbf{x}^T \textbf{Q} \textbf{x} \\
s.t. \quad x_i &\in \left\{0 , 1\right\} \forall i \in \left\{ 1,2, \ldots, N \right\}
\end{aligned}
\label{eq:QUBO}
\end{equation}

In the Ising formualtion, the system energy is given by a Hamiltonian function, as shown in \cref{eq:hamiltonian} \cite{Johnson2011,Boixo2013,Borle2019,Shin2014,Lucas2014,Dwave_gettingstarted, Biswas2016, McGeoch2019,McGeoch2014}.

\begin{equation}
	H \left( \textbf{s} \right) = \sum_{i=1}^{N} h_i s_i + \sum_{i<j} J_{ij} s_i s_j
	\label{eq:hamiltonian}
\end{equation}

In this equation, $\textbf{s} = \left[ s_1 , s_2 , \ldots, s_N \right]$ are Ising spins, which can take values of either -1 or 1. The parameters $h_{i}$ and $J_{ij}$ are qubit biases (self-interaction) and coupling strengths (qubit-qubit interaction) respectively. The summation as defined in \cref{eq:hamiltonian} then yields the Ising Hamiltonian $H$ \cite{Borle2019,Dwave_gettingstarted}. The QA will attempt to find a solution for the Ising spins in $\textbf{s}$ for which the Hamiltonian energy $H$ is minimal.

For this research, the QUBO problem framework is used, because the binary nature of the problem variables provides a convenient foundation to define a truss optimization problem. This will be described in the upcoming section.

\section{Problem Description}
\label{sec:ProblemDescription}

To explore the feasibility of casting a structural optimization problem into a QUBO format, simple two-dimensional truss optimization problems will be investigated. A discrete truss sizing optimization problem can be defined, whereby the cross-sectional area of truss members can be chosen from a predefined discrete set (e.g., a finite number of cross-sectional configurations as provided by the truss supplier). The objective of the optimization is to find the cross-sectional area of each truss member such that the stress in every truss member is as close as possible to the material's limit stress, thereby achieving the optimal use of material and hence minimum weight of the structure.

\begin{figure}[htbp]
	\centering
		\includegraphics[width=0.40\textwidth]{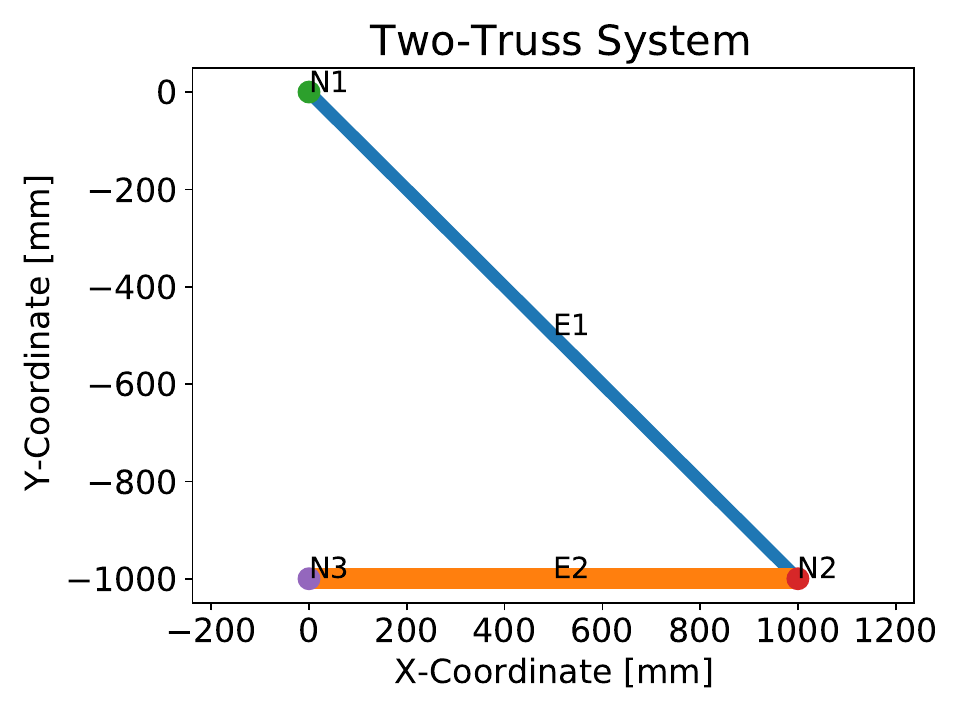}
	\caption{Two-truss system}
	\label{fig:2TrussOriginal}
\end{figure}

\begin{figure}[htbp]
	\centering
		\includegraphics[width=0.40\textwidth]{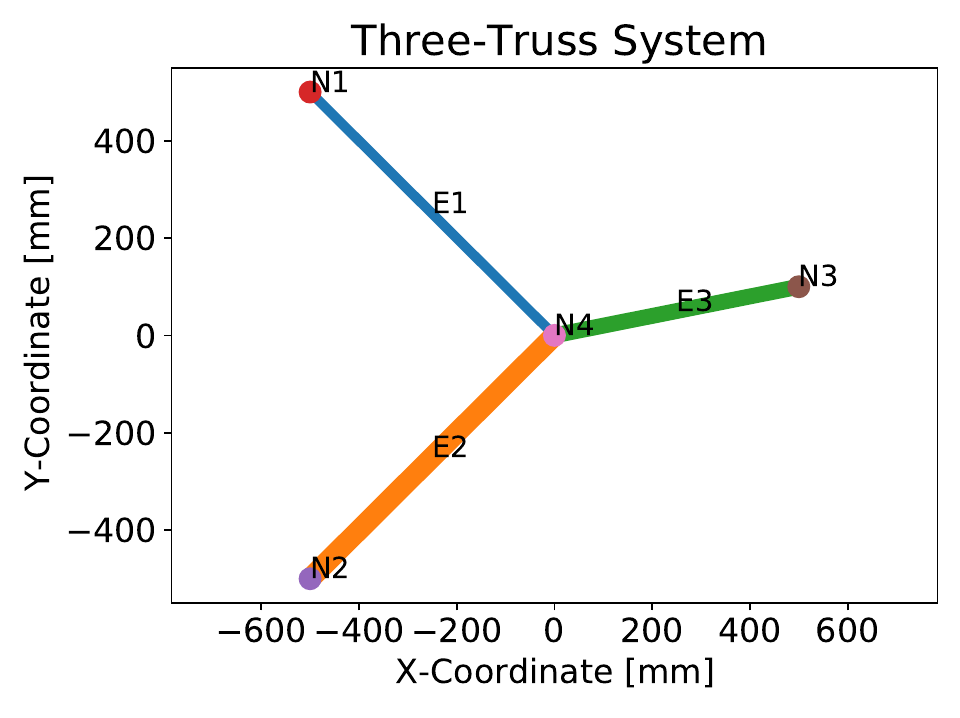}
	\caption{Three-truss system}
	\label{fig:3TrussOriginal}
\end{figure}

\begin{figure}[htbp]
	\centering
		\includegraphics[width=0.40\textwidth]{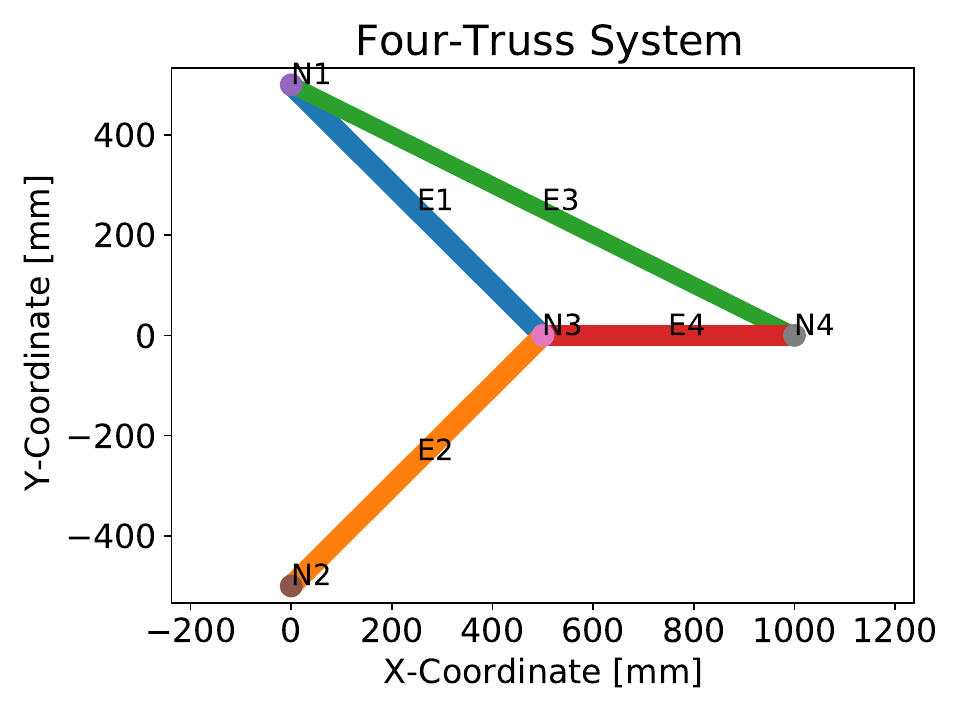}
	\caption{Four-truss system}
	\label{fig:4TrussOriginal}
\end{figure}

The truss systems that are defined for this research incrementally increase in their complexity. In total, three truss systems are defined: a basic two-truss system, a three-truss system, and a slightly more complex four-truss system. The three sample truss systems are shown in \cref{fig:2TrussOriginal,fig:3TrussOriginal,fig:4TrussOriginal}.

The exact definitions of these truss systems, the boundary conditions, and the applied loads are given in \cref{tab:TrussDefinitions}. For each system, the same fictitious material is used, with a Young's modulus of 200~GPa, and a material limit stress of 100~MPa. The material limit stress is assumed to be identical for both compressive and tensile loads. Furthermore, for every truss element, three different allowable choices for the truss cross-sectional area are defined, as shown in \cref{tab:AreaChoices}. \\

\begin{table}[htbp]
\caption{Nodal coordinate definitions, element connectivity, and load and boundary condition definitions. $F_{x/y}$ and $d_{x/y}$ are the prescribed nodal forces and displacements along $x/y$ direction, respectively}
\fontsize{7}{7} \selectfont
\centering
\begin{tabular}{|r|lll|lll|lll|}
\hline
 & \multicolumn{3}{c|}{Two-Truss} & \multicolumn{3}{c|}{Three-Truss} & \multicolumn{3}{c|}{Four-Truss} \\ \hline
\multirow{5}{*}{Nodes} & Node & X {[}mm{]} & Y {[}mm{]} & Node & X {[}mm{]} & Y {[}mm{]} & Node & X {[}mm{]} & Y {[}mm{]} \\
 & N1 & 0 & 0 & N1 & -500 & 500 & N1 & 0 & 500 \\
 & N2 & 1000 & -1000 & N2 & -500 & -500 & N2 & 0 & -500 \\
 & N3 & 0 & -1000 & N3 & 500 & 100 & N3 & 500 & 0 \\
 & - & - & - & N4 & 0 & 0 & N4 & 1000 & 0 \\ \hline
\multirow{5}{*}{Elements} & Element & Start Node & End Node & Element & Start Node & End Node & Element & Start Node & End Node \\
 & E1 & N1 & N2 & E1 & N1 & N4 & E1 & N1 & N3 \\
 & E2 & N2 & N3 & E2 & N2 & N4 & E2 & N2 & N3 \\
 & - & - & - & E3 & N3 & N4 & E3 & N1 & N4 \\
 & - & - & - & - & - & - & E4 & N3 & N4 \\ \hline
\multirow{2}{*}{Load} & Node & $F_{x}$ {[}kN{]} & $F_{y}$ {[}kN{]} & Node & $F_{x}$ {[}kN{]} & $F_{y}$ {[}kN{]} & Node & $F_{x}$ {[}kN{]} & $F_{y}$ {[}kN{]} \\
 & N2 & 0 & -70 & N4 & 0 & -100 & N4 & 0 & -100 \\ \hline
\multirow{4}{*}{BCs} &  Node & $d_{x}$ {[}mm{]} & $d_{y}$ {[}mm{]} & Node & $d_{x}$ {[}mm{]} & $d_{y}$ {[}mm{]} & Node & $d_{x}$ {[}mm{]} & $d_{y}$ {[}mm{]} \\
 & N1 & 0 & 0 & N1 & 0 & 0 & N1 & 0 & 0 \\
 & N3 & 0 & 0 & N2 & 0 & 0 & N2 & 0 & 0 \\
 & - & - & - & N3 & 0 & 0 & - & - & - \\ \hline
\end{tabular}
\label{tab:TrussDefinitions}
\end{table}

\begin{table}[htbp]
\caption{Definition of discrete choices of cross-sectional area for truss system elements.}
\fontsize{7}{7} \selectfont
\centering
\begin{tabular}{|c|ccc|ccc|ccc|}
\hline
         & \multicolumn{3}{c|}{Two-Truss Choices [mm\textsuperscript{2}]} & \multicolumn{3}{c|}{Three-Truss Choices [mm\textsuperscript{2}]} & \multicolumn{3}{c|}{Four-Truss Choices [mm\textsuperscript{2}]} \\ \hline
Elements & Small     & Mid     & Large    & Small     & Mid      & Large     & Small     & Mid      & Large    \\ \hline
E1       & 800       & 900     & 1000     & 400       & 500      & 600       & 2400      & 2500     & 2600     \\ \hline
E2       & 1400      & 1500    & 1600     & 950       & 1050     & 1150      & 2400      & 2500     & 2600     \\ \hline
E3       & -         & -       & -        & 700       & 800      & 900       & 1900      & 2000     & 2100     \\ \hline
E4       & -         & -       & -        & -         & -        & -         & 2400      & 2500     & 2600     \\ \hline
\end{tabular}
\label{tab:AreaChoices}
\end{table}

Having defined the three sample problems, the next step is to start investigating the method by which these discrete truss sizing optimization problems can be cast into a QUBO format. In the upcoming section this process will be described, along with the difficulties and pitfalls that were encountered.

\section{Method}
\subsection{General Concept}
\label{sec:general_concept}

For a truss system consisting of $N$ truss elements, a set of $C$ discrete choices is defined for the truss cross sectional area for every truss element. As such, for every truss element $n$ the set of possible choices can be written as shown in \cref{eq:set_of_A}.

\begin{equation}
A_{n,set}=\left\{A_{n,1} , A_{n,2} , \ldots , A_{n,C} \right\}
\label{eq:set_of_A}
\end{equation}

\noindent To define the cross-sectional area of truss $n$, a set of binary qubit variables is needed that each corresponds to one of the possible choices of cross-sectional area, giving~\cref{eq:set_of_Q}:

\begin{equation}
\begin{aligned}
q_{n,set} = \left\{q_{n,1} , q_{n,2} , \ldots , q_{n,C} \right\} \\ 
\text{With:} \quad q_{n,c} \in \left\{ 0,1 \right\} \ \forall \ c \in \left\{ 1,2,\ldots C \right\}
\end{aligned}
\label{eq:set_of_Q}
\end{equation}

\noindent The total cross-sectional area of truss $n$ can then be defined by summation of the discrete choices and their corresponding binary qubit variable as shown in \cref{eq:TrussArea}.

\begin{equation}
\begin{aligned}
A_n = \sum_{c=1}^{C}{q_{n,c} A_{n,c}}
\end{aligned}
\label{eq:TrussArea}
\end{equation}

In the case that, for truss $n$, only one of the qubits in $q_{n,set}$ is equal to 1, and the others are equal to 0, then this binary variable would correspond directly to a particular choice in cross-sectional area. For a truss system of $N$ truss elements and $C$ discrete choices per truss element, the number of variables needed would become $N \times C$ in total. Together, they form the solution vector of the problem.  For example, for the two-truss problem, the solution vector would be defined as: 




\begin{equation}
\begin{aligned}
\textbf{q} &= [q_{1,1}, \quad q_{1,2}, \quad q_{1,3}, \quad q_{2,1}, \quad q_{2,2}, \quad q_{2,3}] \\
\end{aligned}
\label{eq:example_solution_vector}
\end{equation}

To select the mid-sized choice for each truss element in the two-truss problem, the solution vector would evaluate to:

\begin{equation}
\begin{aligned}
\textbf{q} &= [0, \quad 1, \quad 0, \quad 0, \quad 1, \quad 0] \\
\end{aligned}
\label{eq:example_solution_vector_filled}
\end{equation}



It may be noted at this point that it is technically possible to select multiple cross-sectional areas for a single truss element. This would occur when a truss element has more than one associated binary variable set to a value of $1$. Per \cref{eq:TrussArea}, this would mean that the area of the truss element becomes the summation of multiple available choices. However, in this research the goal is to make a single distinct choice from the available set of choices. Solutions where exactly one cross-sectional area is selected per truss element will be considered valid solutions. Any other potential solution, selecting either none, or multiple cross-sectional areas per truss element will be considered invalid. 

With the truss cross-sectional area defined in terms of qubit variables as shown above, a symbolic solution process for the truss optimization problems has been conceived, such that the objective function would eventually be expressed as a QUBO function of these qubit variables. We would then use QA to find the solutions of these qubit variables which would minimize the objective function. The following steps summarize this symbolic solution process:

\begin{enumerate}
	\item Using the expression for the truss cross-sectional area, the element stiffness matrices of the truss members can also be written in terms of the qubit variables. 
	\item Using the FEM assembly procedure, the symbolic global stiffness matrix of the entire truss structure can be assembled from each of the element stiffness matrices. 
	\item Proceeding as normal with the FEM analysis, boundary conditions must be taken into account, and a vector of applied loads must be known. By inverting the symbolic global stiffness matrix, and multiplying this inverse matrix with the load vector, a symbolic vector of nodal displacements can be obtained. \label{step_invert}
	\item Using the symbolic vector of nodal displacements, and the known initial length of every truss element, symbolic expressions for the strains of truss elements can be set up.  
	\item By multiplying the symbolic expressions of the truss strains with the Young's modulus, symbolic expressions for the truss stresses are obtained.  
    \item The symbolic expressions for the truss stresses will be used to construct an objective function for which the minimum solution encodes the optimal choice of cross-sectional area for every truss element in the structure. 
    \item The symbolic objective function will be transformed into a QUBO format, then sent to D-Wave to find the minimum solution. 
\end{enumerate}

\subsection{Symbolic Finite Element Method}
\label{sec:SFEM}


\subsubsection{Finding Expressions for Nodal Displacement}
\label{sec:SFEM_disp}

In the displacement-based linear finite element method for truss structures, the nodal Degrees of Freedom (DoFs) are the displacements. The stiffness matrix of each truss element can be defined by its nodal coordinates, cross-sectional area, and the Young's modulus of the material. These element stiffness matrices are then assembled according to the element's connectivity matrix to form a global system equation, typically as shown in \cref{eq:FEM}. 

\begin{equation}
	\textbf{K} \, \textbf{u} = \textbf{f}
	\label{eq:FEM}
\end{equation}

The matrix $\textbf{K} $ is known, and represents the global stiffness matrix of the finite-element structure. The vector $\textbf{f}$ is also known, as this vector defines the forces which are applied to the structure. The goal for the linear finite-element problem is to find the solution vector $\textbf{u}$, which contains the displacement of every node in the structure. By knowing the displacements of all nodes in the structure, other metrics such as the element strain and stress can also be calculated.

 As the element stiffness matrix depends on the cross-sectional area, which is now represented symbolically in \cref{eq:TrussArea}, the stiffness matrix of truss element $n$ will then be a function of the associated qubit variables in \cref{eq:set_of_Q}. The assembled global stiffness matrix will then be a function of all the qubit variables of this structure, i.e., $\textbf{K}$ will be a function of the vector $\textbf{q}$ as shown in \cref{eq:example_solution_vector} for the two-truss example. A script has been written that can set up the truss finite-element system equation symbolically: 
 \begin{equation}
	\textbf{K}(\textbf{q}) \, \textbf{u} = \textbf{f}
	\label{eq:qFEM}
\end{equation}
 The above symbolic system equation has been implemented and solved in both Python and Matlab. Based on our experience, the Matlab backslash operator seems to solve for the solution $\textbf{u}$ faster than Python Sympy. Hence, it is used here to obtain the symbolic nodal displacement vector: 
\begin{equation}
	\textbf{u}(\textbf{q}) = \textbf{K}(\textbf{q}) \backslash \textbf{f}
	\label{eq:FEM_matlab}
\end{equation} 
 Having obtained the symbolic expressions for the nodal displacements, it became clear why this had been troublesome to calculate. The expressions for the nodal displacements are extremely long, even for such simple finite-element problems. This constitutes a bottleneck of the proposed approach, on which we will discuss later. Nevertheless, once we have obtained these expressions, the symbolic finite-element process can be continued.

\subsubsection{Finding Expressions for Strain}
\label{sec:SFEM_strain}
Engineering strain is a common choice of strain measure for truss elements:
\begin{equation}
\epsilon = \frac{L_{\mathit{disp}} - L_{0}}{L_{0}}
\label{eq:strain_init}
\end{equation} 
where the original length of the truss element is denoted as $L_{0}$, and the deformed length of the truss element as $L_{\mathit{disp}}$. However, when this strain was implemented, it was found that this leads to a symbolic expression for the truss strain that contains many absolute functions. The appearance of these absolute functions was problematic as they prevented the symbolic expression from being written purely as a sum of quadratic polynomial terms, which is eventually necessary for the QUBO problem formulation. 

To circumvent the appearance of the absolute functions in the symbolic expressions for the truss strain, the Green-Lagrange strain formulation was implemented, as given in \cref{eq:strain_GL}. Under the infinitesimal deformation assumption, when $L_{0} \approx L_{\mathit{disp}}$, \cref{eq:strain_init} and \cref{eq:strain_GL} are equivalent.

\begin{equation}
\epsilon = \frac{1}{2}\left(\frac{L_{\mathit{disp}}^2}{L_{0}^2} - 1\right)
\label{eq:strain_GL}
\end{equation}

With the Green-Lagrange strain implementation it was seen that the absolute functions no longer appeared in the symbolic truss strain expressions. Thus, the symbolic expressions for the strain of every truss element in the sample truss sizing optimization problems could be found. In turn, these can then be used to find expressions for the truss element stresses.

\subsubsection{Expressions for Stress}
\label{sec:SFEM_stress}
Expressions for truss stresses follow from \cref{eq:truss_stress}, in which the material Young's modulus is denoted by $E$.

\begin{equation}
\sigma = E \, \epsilon
\label{eq:truss_stress}
\end{equation}



To give an example of the symbolic expressions for the truss element stresses, \cref{eq:truss_stress1,eq:truss_stress2,eq:truss_stress3} give the stress in truss element 1 for the two-truss problem. Note, the expression for the stress in the truss element takes a fractional form. 

\begin{equation}
\begin{aligned}
\sigma_{1} =& \frac{N_1}{D_1} \\
\end{aligned}
\label{eq:truss_stress1}
\end{equation}

\noindent With:

\noindent
\begin{center}
\footnotesize
    \begin{minipage}[c]{0.5\textwidth}
        \begin{equation}
        \begin{aligned}
        N_1 =& 7.4046706\times 10^{29}\times q_{1,1} \\
        &+9.3715363\times 10^{29}\times q_{1,2}\\
        &+1.1569798\times 10^{30}\times q_{1,3}\\
        &+9.0707215\times 10^{30}\times q_{2,1}\\
        &+1.0412818\times 10^{31}\times q_{2,2}\\
        &+1.1847473\times 10^{31}\times q_{2,3}\\
        &+1.6660509\times 10^{30}\times q_{1,1}\times q_{1,2}\\
        &+1.8511677\times 10^{30}\times q_{1,1}\times q_{1,3}\\
        &+2.0825636\times 10^{30}\times q_{1,2}\times q_{1,3}\\
        &+1.4664165\times 10^{34}\times q_{1,1}\times q_{2,1}\\
        &+1.6833582\times 10^{34}\times q_{1,1}\times q_{2,2}\\
        &+1.6497186\times 10^{34}\times q_{1,2}\times q_{2,1}\\
        &+1.9152597\times 10^{34}\times q_{1,1}\times q_{2,3}\\
        &+1.893778\times 10^{34}\times q_{1,2}\times q_{2,2}\\
        &+1.8330206\times 10^{34}\times q_{1,3}\times q_{2,1}\\
        &+2.1546671\times 10^{34}\times q_{1,2}\times q_{2,3}\\
        &+2.1041978\times 10^{34}\times q_{1,3}\times q_{2,2}\\
        &+2.3940746\times 10^{34}\times q_{1,3}\times q_{2,3}\\
        &+1.943726\times 10^{31}\times q_{2,1}\times q_{2,2}\\
        &+2.0733078\times 10^{31}\times q_{2,1}\times q_{2,3}\\
        &+2.2214012\times 10^{31}\times q_{2,2}\times q_{2,3}\\
        &+3.1415357\times 10^{34}\times q_{1,1}\times q_{2,1}\times q_{2,2}\\
        &+3.3509714\times 10^{34}\times q_{1,1}\times q_{2,1}\times q_{2,3}\\
        &+3.5342277\times 10^{34}\times q_{1,2}\times q_{2,1}\times q_{2,2}\\
        &+3.5903265\times 10^{34}\times q_{1,1}\times q_{2,2}\times q_{2,3}\\
        &+3.7698428\times 10^{34}\times q_{1,2}\times q_{2,1}\times q_{2,3}\\
        &+3.9269196\times 10^{34}\times q_{1,3}\times q_{2,1}\times q_{2,2}\\
        &+4.0391173\times 10^{34}\times q_{1,2}\times q_{2,2}\times q_{2,3}\\
        &+4.1887143\times 10^{34}\times q_{1,3}\times q_{2,1}\times q_{2,3}\\
        &+4.4879081\times 10^{34}\times q_{1,3}\times q_{2,2}\times q_{2,3}\\
        \end{aligned}
        \label{eq:truss_stress2}
        \end{equation}    
    \end{minipage}%
    \begin{minipage}[c]{0.5\textwidth}
        \begin{equation}
        \begin{aligned}
        D_1 =& 1.1847473\times 10^{32}\times q_{1,1}\times q_{2,1}\\
        &+1.3600415\times 10^{32}\times q_{1,1}\times q_{2,2}\\
        &+1.4994458\times 10^{32}\times q_{1,2}\times q_{2,1}\\
        &+1.547425\times 10^{32}\times q_{1,1}\times q_{2,3}\\
        &+1.7213026\times 10^{32}\times q_{1,2}\times q_{2,2}\\
        &+1.8511677\times 10^{32}\times q_{1,3}\times q_{2,1}\\
        &+1.9584598\times 10^{32}\times q_{1,2}\times q_{2,3}\\
        &+2.1250649\times 10^{32}\times q_{1,3}\times q_{2,2}\\
        &+2.4178516\times 10^{32}\times q_{1,3}\times q_{2,3}\\
        &+2.6656814\times 10^{32}\times q_{1,1}\times q_{1,2}\times q_{2,1}\\
        &+3.0600935\times 10^{32}\times q_{1,1}\times q_{1,2}\times q_{2,2}\\
        &+2.9618683\times 10^{32}\times q_{1,1}\times q_{1,3}\times q_{2,1}\\
        &+3.4817064\times 10^{32}\times q_{1,1}\times q_{1,2}\times q_{2,3}\\
        &+3.4001039\times 10^{32}\times q_{1,1}\times q_{1,3}\times q_{2,2}\\
        &+3.3321018\times 10^{32}\times q_{1,2}\times q_{1,3}\times q_{2,1}\\
        &+3.8685626\times 10^{32}\times q_{1,1}\times q_{1,3}\times q_{2,3}\\
        &+3.8251169\times 10^{32}\times q_{1,2}\times q_{1,3}\times q_{2,2}\\
        &+4.352133\times 10^{32}\times q_{1,2}\times q_{1,3}\times q_{2,3}\\
        &+2.5387442\times 10^{32}\times q_{1,1}\times q_{2,1}\times q_{2,2}\\
        &+2.7079938\times 10^{32}\times q_{1,1}\times q_{2,1}\times q_{2,3}\\
        &+3.2130982\times 10^{32}\times q_{1,2}\times q_{2,1}\times q_{2,2}\\
        &+2.901422\times 10^{32}\times q_{1,1}\times q_{2,2}\times q_{2,3}\\
        &+3.4273047\times 10^{32}\times q_{1,2}\times q_{2,1}\times q_{2,3}\\
        &+3.9667878\times 10^{32}\times q_{1,3}\times q_{2,1}\times q_{2,2}\\
        &+3.6721122\times 10^{32}\times q_{1,2}\times q_{2,2}\times q_{2,3}\\
        &+4.2312404\times 10^{32}\times q_{1,3}\times q_{2,1}\times q_{2,3}\\
        &+4.5334718\times 10^{32}\times q_{1,3}\times q_{2,2}\times q_{2,3}\\
        &+5.7121745\times 10^{32}\times q_{1,1}\times q_{1,2}\times q_{2,1}\times q_{2,2}\\
        &+6.0929861\times 10^{32}\times q_{1,1}\times q_{1,2}\times q_{2,1}\times q_{2,3}\\
        &+6.3468606\times 10^{32}\times q_{1,1}\times q_{1,3}\times q_{2,1}\times q_{2,2}\\
        &+6.5281994\times 10^{32}\times q_{1,1}\times q_{1,2}\times q_{2,2}\times q_{2,3}\\
        &+6.7699846\times 10^{32}\times q_{1,1}\times q_{1,3}\times q_{2,1}\times q_{2,3}\\
        &+7.1402181\times 10^{32}\times q_{1,2}\times q_{1,3}\times q_{2,1}\times q_{2,2}\\
        &+7.2535549\times 10^{32}\times q_{1,1}\times q_{1,3}\times q_{2,2}\times q_{2,3}\\
        &+7.6162327\times 10^{32}\times q_{1,2}\times q_{1,3}\times q_{2,1}\times q_{2,3}\\
        &+8.1602493\times 10^{32}\times q_{1,2}\times q_{1,3}\times q_{2,2}\times q_{2,3}\\
        \end{aligned}
        \label{eq:truss_stress3}
        \end{equation}
    \end{minipage}
\end{center}
For reference, the symbolic expressions for the truss stresses of each of the sample problems are available online \cite{DataverseAnalyticalQUBO}.



Now that a method has been implemented that can write the stresses in terms of qubit variables, the next step is to set up an objective function that, when minimized, should yield the most optimal choice of cross-sectional area for every truss element in the structure. However, this process uncovered another challenge, which is discussed in the upcoming section.

\subsection{Development of an Objective Function}
\label{sec:SetupObjective}

\subsubsection{Fractional Objective Function}

Using the expression for the stress in a truss element, an objective function can be set up describing the difference between the truss element stress and the maximum limit stress allowed by the material properties. If such a difference is minimized, then the truss element will be as close as possible to the material failure stress, meaning the weight of the truss is implicitly minimized.

When setting up the objective function, it is important to consider that the truss element stress evaluates to a negative number for truss elements under compression. However, because it is not known beforehand which truss elements will experience compressive or tensile stresses, this poses an issue. It is not possible to apply an absolute function to the truss element stress, as such a mathematical function is incompatible with QUBO problem formulations. As an alternative, the expression for the truss element stress can be squared to ensure that it always evaluates to a positive number. To keep consistent units, this also means that the maximum allowable stress is squared. A minimization problem is obtained by squaring the difference between the squared material limit stress and the squared truss element stress. As such, for a single truss element $n$, a minimization objective function can be defined as shown in \cref{eq:target2}. 
 
\begin{equation}
F_{n} = \left(\sigma_{\mathit{limit}}^2 - \sigma_{n}^2 \right)^2
\label{eq:target2}
\end{equation}

With \cref{eq:target2}, the minimum solution should encode the choice of truss cross-sectional area that minimizes the absolute difference between the material limit stress and the truss stress. To set up an objective function that describes the entire system of truss elements, the summation is taken of the objective functions of every individual truss element, which leads to \cref{eq:target3}. 

\begin{equation}
F = \sum_{n=1}^{N}{F_{n}} = \sum_{n=1}^{N}{\left(\sigma_{\mathit{limit}}^2 - \sigma_{n}^2 \right)^2}
\label{eq:target3}
\end{equation}

\begin{figure}[h!]
	\centering
		\includegraphics[height=6.5cm]{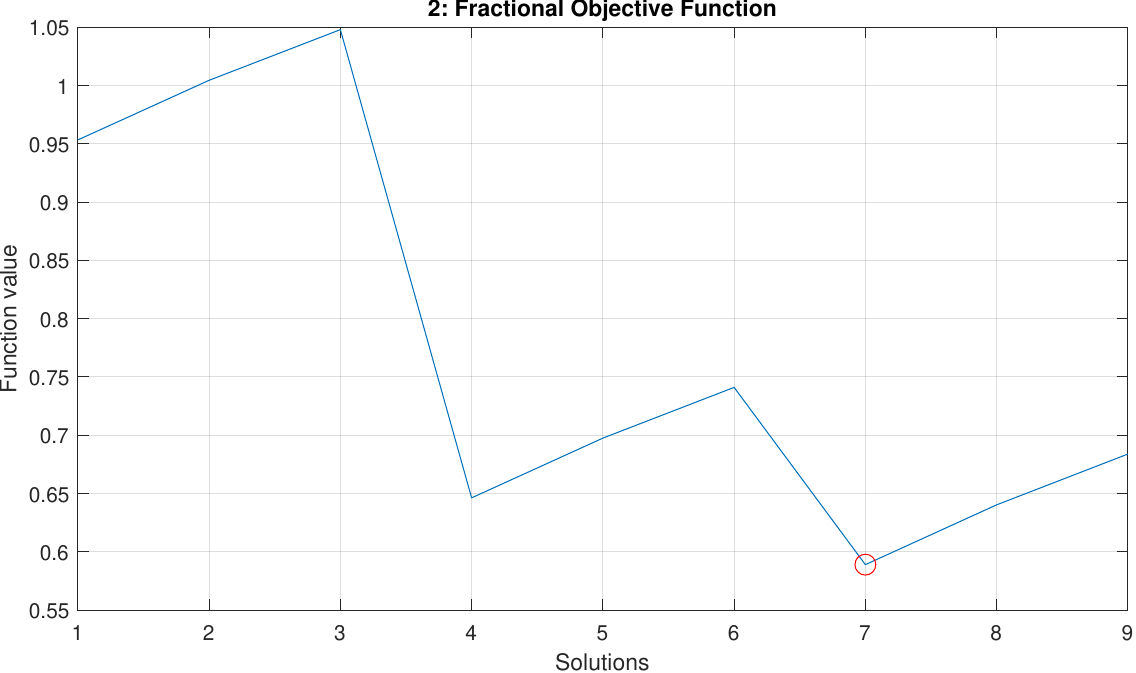}
	\caption{Fractional objective function for the two-truss problem. Global minimum is found to be solution number 7, corresponding to $\left[ 0,0,1,1,0,0 \right]$.}
	\label{fig:2Fraction_Obj}
\end{figure}

\begin{figure}[h!]
	\centering
		\includegraphics[height=6.5cm]{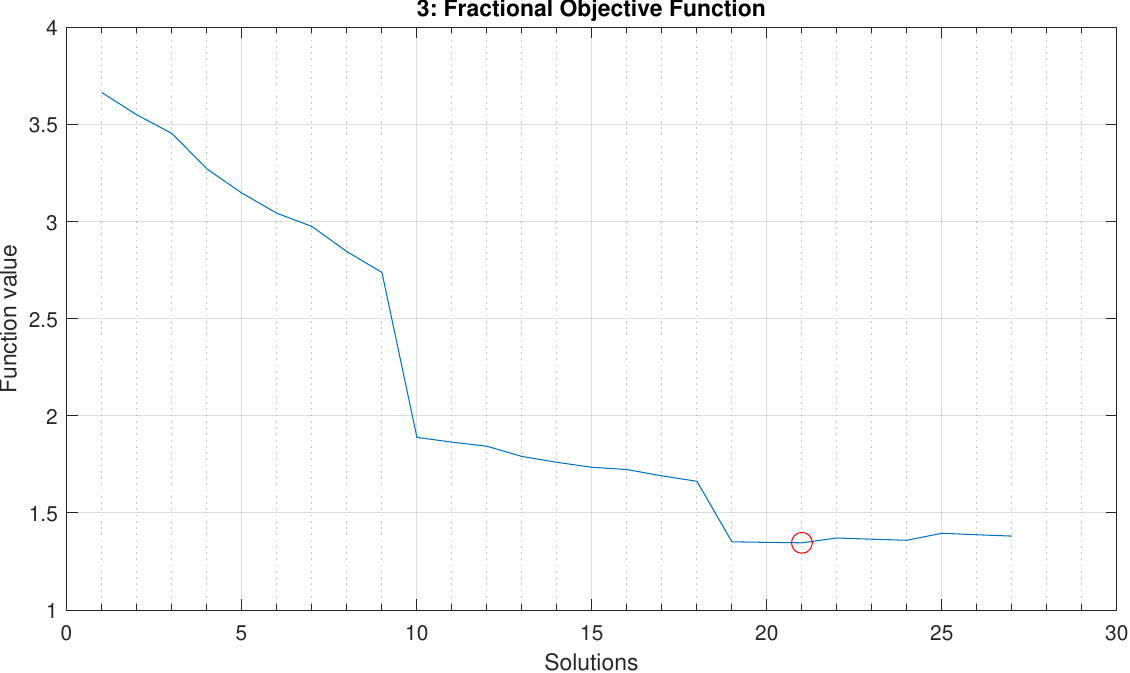}
	\caption{Fractional objective function for the three-truss problem. Global minimum is found to be solution number 21, corresponding to $\left[ 0,0,1,1,0,0,0,0,1 \right]$.}
	\label{fig:3Fraction_Obj}
\end{figure}

\begin{figure}[h!]
	\centering
		\includegraphics[height=6.5cm]{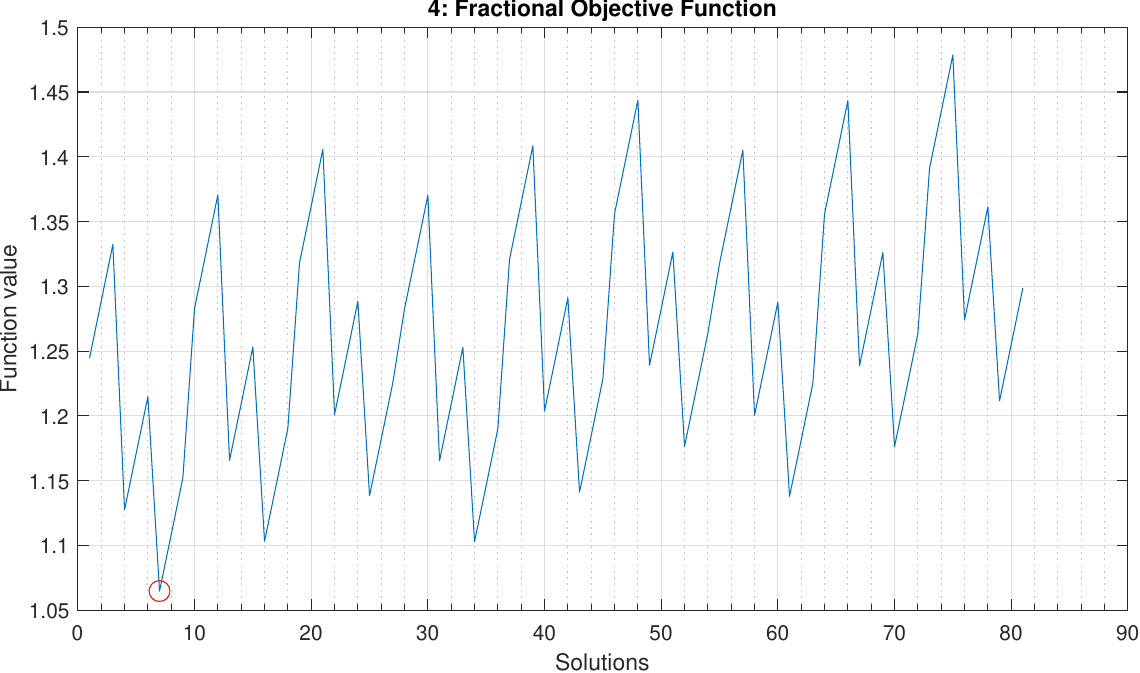}
	\caption{Fractional objective function for the four-truss problem. Global minimum is found to be solution number 7, corresponding to $\left[ 1,0,0,1,0,0,0,0,1,1,0,0 \right]$.}
	\label{fig:4Fraction_Obj}
\end{figure}

With \cref{eq:target3} a general method is obtained that can be used to set up an objective function for each of the sample problems. It is important to note that, because the truss element stresses $\sigma_n$ are fractional in nature, as was shown in \cref{eq:truss_stress1,eq:truss_stress2,eq:truss_stress3}, the objective function from \cref{eq:target3} will also have a fractional form. Hence, the resulting objective function is referred to as the fractional objective function. To test whether this method of setting up an objective function works as expected, the objective functions are set up for each of three sample problems, and are evaluated by a brute-force analysis, i.e., the objective function is evaluated at every possible solution. The results for the three sample problems are shown in \cref{fig:2Fraction_Obj,fig:3Fraction_Obj,fig:4Fraction_Obj}, and were produced using the Matlab code which is available online \cite{DataverseAnalyticalQUBO}. The following global minimum solutions were found:

\begin{itemize}
    \item Two-truss problem: minimum at solution number 7, $\left[ 0,0,1,1,0,0 \right]$
    
    \item Three-truss problem: minimum at solution number 21, $\left[ 0,0,1,1,0,0,0,0,1 \right]$

    \item Four-truss problem: minimum at solution number 7, $\left[ 1,0,0,1,0,0,0,0,1,1,0,0 \right]$
\end{itemize}

The above brute-force results will be used as references to benchmark the performance of QA.   One may be tempted to directly send the objective function in \cref{eq:target3} to QA and see how it performs. However, one challenging aspect remains: the function has a fractional form. Fractional objective functions are incompatible with QUBO problem formulations, and cannot be directly solved by the QA. As such, in the next section, a method is investigated by which a potential non-fractional objective function can be set up that approximates the original fractional one.

\subsubsection{Non-Fractional Objective Function}
In a QUBO-based objective function, only a summation of linear and quadratic terms is allowed. Therefore, in this section a potential method is investigated to formulate a non-fractional objective function for the discrete truss sizing optimization problem. 

The reserve factor of a truss element gives a measure of how close the truss element is to material failure. The reserve factor for a truss element $n$ is calculated via \cref{eq:truss_RF2}.

\begin{equation}
RF_{n} = \frac{\sigma_{\mathit{limit}}}{\vert\sigma_{n}\vert}
\label{eq:truss_RF2}
\end{equation}

To mitigate the absolute function for compatibility with QUBO, the material limit stress $\sigma_{limit}$ and truss element stress $\sigma_n$ can be squared. By doing so, a squared reserve factor can be calculated, according to \cref{eq:RF_sq1}.

\begin{equation}
RF^{2}_{n} = \frac{\sigma_{\mathit{limit}}^2}{\sigma^{2}_{n}}
\label{eq:RF_sq1}
\end{equation}

Using \cref{eq:RF_sq1} the theoretical optimal value of the squared reserve factor shall be 1 if the cross-sectional area can be varied arbitrarily. Given that the squared reserve factor is written as a fraction, this means that the numerator and the denominator should be equal to each other in the optimal case. In this case the numerator minus the denominator should also yield a result of zero. However, assuming that the currently known symbolic expression for the truss reserve factor actually describes a sub-optimal case, this will result in an error term. An optimization problem can then be set up for which the goal is to minimize this error, by simply squaring the expression. This ensures that the optimum solution will be the one where the squared error is closest to zero, giving an objective function for the minimization. The steps to this process are shown in \cref{eq:RF_opt}.

\begin{equation}
\begin{aligned}
\text{If: }& RF^{2}_{n} = \frac{\sigma_{\mathit{limit}}^2}{\sigma^{2}_{n}} \quad \text{and} \quad \sigma^{2}_{n} = \frac{\sigma^{2}_{N,n}}{\sigma^{2}_{D,n}}\\
\text{Rewriting: }&   RF^{2}_{n} = \frac{\sigma_{\mathit{limit}}^2  \sigma^{2}_{D,n} }{\sigma^{2}_{N,n}} = \frac{N_n}{D_n} \\
\text{When: }& RF^{2}_{n} = \frac{N_n}{D_n} = 1 \\
\text{Then: }& N_n=D_n \\
\text{Optimally: }& N_n-D_n = 0 \\
\text{Sub-optimally: }&   N_n-D_n = \epsilon \\
\text{Optimizing: }& \min{\left(\epsilon^2\right)} \Rightarrow \min{\left(\left(N_n-D_n\right)^{2}\right)} \\
\end{aligned}
\label{eq:RF_opt}
\end{equation}

Considering the above reasoning, the non-fractional objective function for a truss element $n$ is written in \cref{eq:target_sq1}.

\begin{equation}
F_n = (N_n - D_n)^2 = \left( \sigma_{\mathit{limit}}^2 \sigma^{2}_{D,n}  -  \sigma^{2}_{N,n} \right)^{2}
\label{eq:target_sq1}
\end{equation}

The objective function for the complete truss system is then found by taking the sum for all truss elements, as given in \cref{eq:target_sq2}. 

\begin{equation}
F = \sum_{n=1}^{N}{F_n}  = \sum_{n=1}^{N}{\left( \sigma_{\mathit{limit}}^2 \sigma^{2}_{D,n}  -  \sigma^{2}_{N,n} \right)^{2}}
\label{eq:target_sq2}
\end{equation}

\begin{figure}[htbp]
	\centering
		\includegraphics[height=6.5cm]{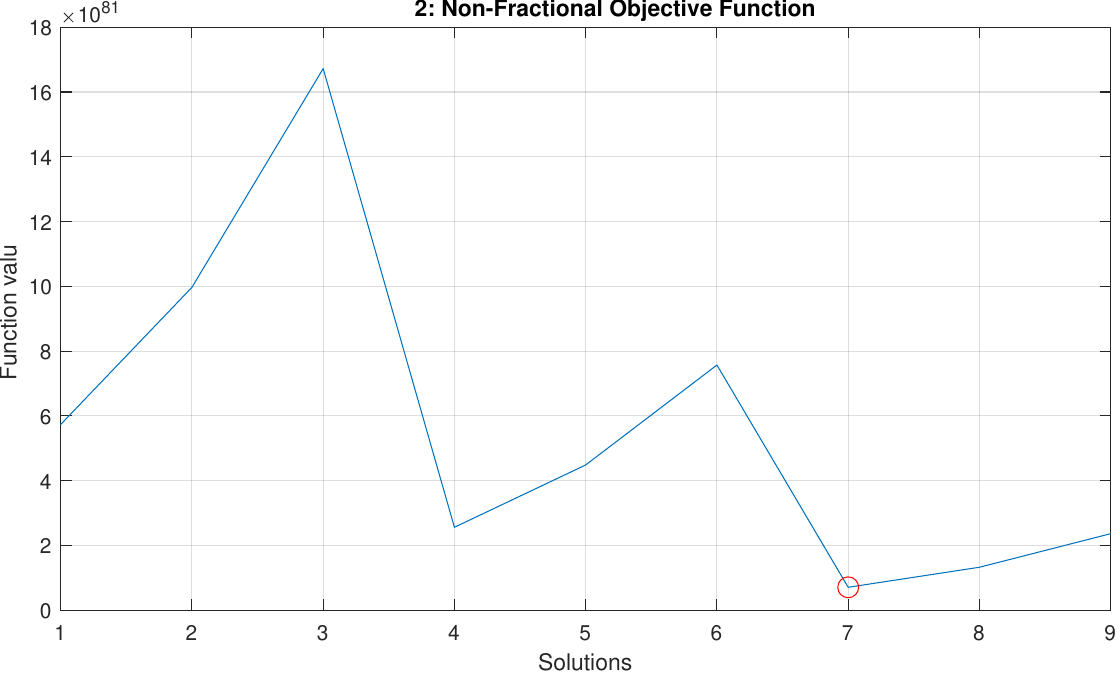}
	\caption{Non-fractional objective function for the two-truss problem. Global minimum is found to be solution number 7, corresponding to $\left[ 0,0,1,1,0,0 \right]$.}
	\label{fig:2Improved}
\end{figure}

\begin{figure}[htbp]
	\centering
		\includegraphics[height=6.5cm]{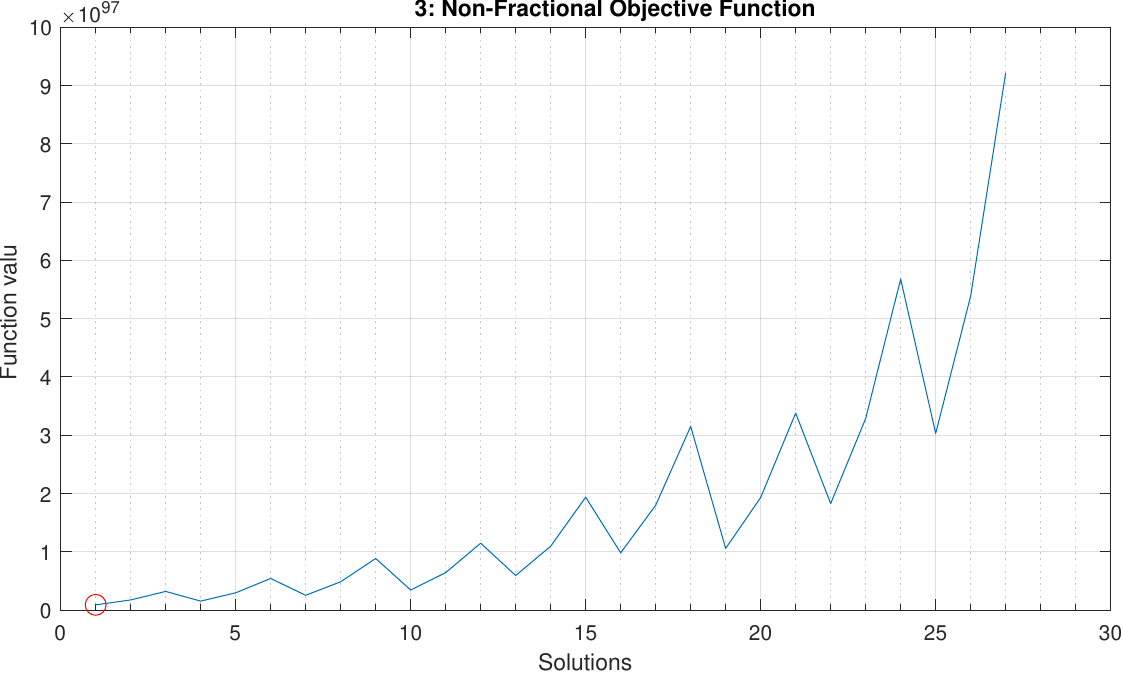}
	\caption{Non-fractional objective function for the three-truss problem. Global minimum is found to be solution number 1, corresponding to $\left[1,0,0,1,0,0,1,0,0\right]$.}
	\label{fig:3Improved}
\end{figure}

\begin{figure}[htbp]
	\centering
		\includegraphics[height=6.5cm]{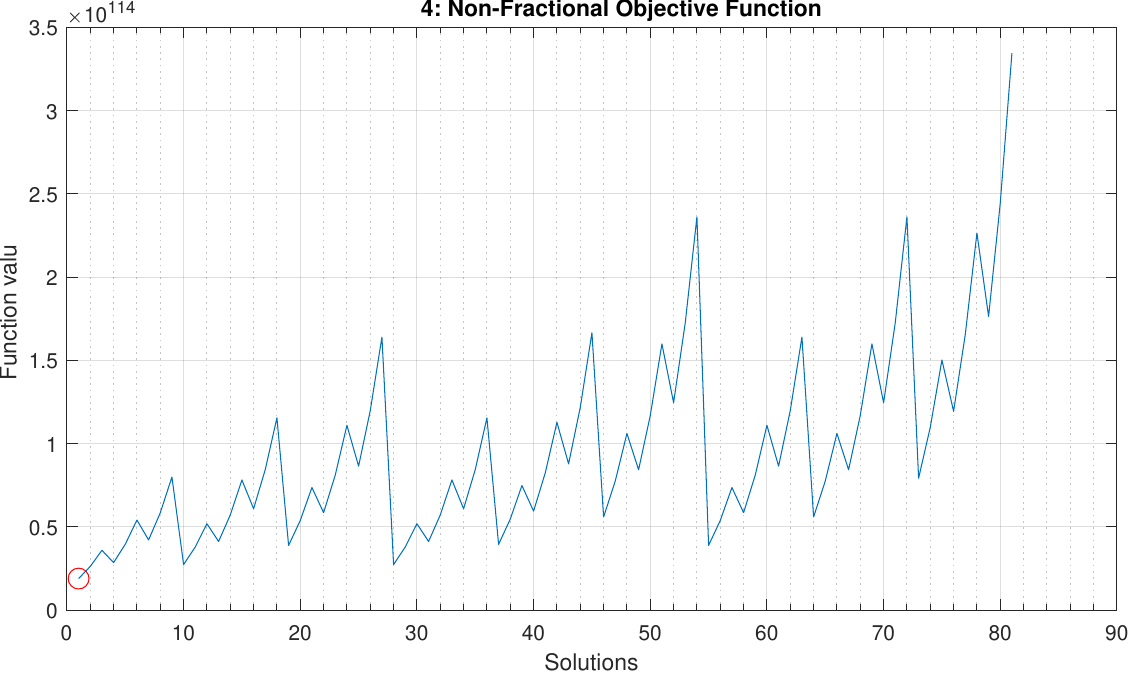}
	\caption{Non-fractional objective function for the four-truss problem. Global minimum is found to be solution number 1, corresponding to $\left[1,0,0,1,0,0,1,0,0,1,0,0\right]$.}
	\label{fig:4Improved}
\end{figure}

This reformulated non-fractional objective function is again tested with the sample problems, using the brute-force analysis method \cite{DataverseAnalyticalQUBO}, to find out if the minimum solutions remain the same as those of the original fractional objective function. The results of these analyses are shown in \cref{fig:2Improved,fig:3Improved,fig:4Improved}, with the global minimum solutions indicated with a small red circle. The following global optimum solutions were obtained:

\begin{itemize}
    \item Two-truss problem: minimum at solution number 7, $\left[ 0,0,1,1,0,0 \right]$
    
    \item Three-truss problem: minimum at solution number 1, $\left[1,0,0,1,0,0,1,0,0\right]$

    \item Four-truss problem: minimum at solution number 1, $\left[1,0,0,1,0,0,1,0,0,1,0,0\right]$
\end{itemize}

Using the non-fractional objective function, the expected solution for the two-truss problem is returned. However, for the three- and four-truss problems, the results are however not those expected. Therefore this method is presumed to be flawed. In the method, the difference between $N_n$ and $D_n$ is considered, rather than considering the ratio $N_n/D_n$ which defines the squared reserve factor. A solution that minimizes the difference between $N_n$ and $D_n$ may not necessarily be the same solution for which the ratio between $N_n$ and $D_n$ is closest to a value of 1. Hence, this may be the source of the unexpected behavior produced by this objective function method. In the next section, an alternative method is discussed, with which a fractional objective function can still be made compatible with the QA.

\subsection{Iterative non-fractional approximations to the fractional objective function}

\citeauthor{Ajagekar2019} developed a method with which the optimum solution to fractional objective functions can be found, by iteratively solving an adaptive non-fractional function \cite{Ajagekar2019}. The iterative scheme described by the authors is presented in \cref{eq:Ajagekar_iter}.

\noindent
\makebox[\linewidth][c]{
\begin{minipage}{\linewidth}
	\begin{equation}
	\begin{aligned}%
    \text{With :}& \: F = \frac{N (\mathbf{q}) }{D (\mathbf{q}) }\\
	0:& \: \text{Initialize variables}\\
	&\qquad \mathit{iter} = 0 \\
	&\qquad \lambda = 0 \\
	&\qquad \mathit{obj} = \infty \\
	&\qquad \delta = 10^{-6} \\
	1:& \: \mathbf{while} \left| \lambda - obj \right| > \delta \\
	2:&\qquad \mathit{iter} = \mathit{iter} + 1 \\
	3:&\qquad F_{nf}\left(\mathbf{q}\right) = N \left(\mathbf{q}\right) - \lambda D\left(\mathbf{q}\right) \\
	4:&\qquad \text{find} \:\:\: \hat{\mathbf{q}} \:\:\: \text{s.t.} \:\:\: \min{\left(F_{nf}\left(\hat{\mathbf{q}}\right)\right)} \\
	5:&\qquad \mathit{obj} = \lambda \\
	6:&\qquad \lambda = F\left(\hat{\mathbf{q}}\right)\\ 
	\end{aligned}
	\label{eq:Ajagekar_iter}
	\end{equation}
\end{minipage}}

In this scheme the fractional objective function $F\left(\mathbf{q}\right)$, written in terms of the binary problem variables $\mathbf{q}$, is rewritten into a non-fractional form. This happens in step 3 of the scheme, producing the non-fractional objective $F_{nf}\left(\mathbf{q}\right)$. If this non-fractional function can be written in a QUBO form, then the partially optimal solution $\hat{\mathbf{q}}$ can be found using the QA, in step 4 of the iterative scheme. In step 6, the objective function value of the original fractional function is evaluated using the partially optimal solution that was found to the non-fractional function. When the difference between the objective function values found in consecutive iterations is less than $\delta$ the while-loop is broken and the analysis has converged. Ultimately, the final solution that is found for $\hat{\mathbf{q}}$ will be the minimum solution to the original fractional objective function. 

This iterative method for finding the optimal solution to fractional objective functions was implemented in the Python scripts for the discrete truss sizing optimization problems~\cite{DataverseAnalyticalQUBO}. However, the method was slightly extended, to include a user-defined maximum allowed number of iterations. This was added as an additional condition to while-loop in step 1 of \cref{eq:Ajagekar_iter}. In other words, defining step 1 as:

\begin{equation}
    \mathbf{while} \left| \lambda - obj \right| > \delta \quad AND \quad \mathit{iter} \leq \mathit{iter}_{max}
\end{equation}

\noindent Setting a maximum number of iterations will help to prevent wasting excessive quantum computational time in the case that the analysis has difficulty converging on an optimal solution.

\subsection{Objective Function Processing to Yield a QUBO Problem}
With the iterative scheme that was defined by \citeauthor{Ajagekar2019} in \cite{Ajagekar2019}, the fractional objective function for the discrete truss sizing optimization problem can be solved by iteratively finding the solution to an adaptive non-fractional function. Once this non-fractional form is determined, there are a number of processing and simplification steps that can be taken to finally yield a QUBO problem. By reducing the complexity of the objective function, the QA might be able to more easily identify the global optimum solution. The processing steps will allow the user to fine-tune the performance of the QA more easily, and aid in finding valid solutions to the discrete truss sizing optimization problem. In the following sections the simplifications and processing steps are discussed.

\subsubsection{High-Order Truncation}
The first simplification to the rewritten non-fractional objective function is to truncate excessively high-order terms. Specifically, for a system with $N$ truss elements, any term in the objective function that is above $N^{\text{th}}$order can be truncated. These terms do not contribute any useful information about valid solutions to the optimization problem. Since, for a valid solution to the truss optimization problem, only $N$ number of qubits are expected to end in a 1 state, while all other qubits should end in a 0 state. Therefore, all terms in the objective function that are above $N^{\text{th}}$ order never contribute to valid solutions. These terms can therefore safely be truncated in order to simplify the overall objective function, while having no influence on valid problem solutions. 

The objective functions generally contain every possible unique multiplication between the qubit variables. Therefore, truncating the excessively high-order terms from the objective function can have a very large impact on the overall complexity of the objective function. This was investigated with a simple Excel sheet, which is made available online \cite{DataverseAnalyticalQUBO}. For example, in the two-truss problem, a total of six qubit variables are used. In total, this gives 63 different unique multiplications of qubit variables from first to sixth order. For this problem, all terms above second order only contribute information to invalid problem solutions, as for valid solutions it is expected that exactly two qubit variables will be given a value of $1$. As such, terms above second order can all safely be removed from the objective function. After truncating the terms above second order, only 21 total first- and second-order terms remain in the objective function. This means that the overall number of terms in the objective function is reduced by 66\%. In a similar vein, for the three-truss problem, this truncation reduces the number of terms in the objective function by roughly 75\%. For the four-truss problem, the reduction is approximately 80\%. It is expected that such a significant reduction in the complexity of the objective function will benefit the QA, allowing it to more easily find valid global minimum solutions to the truss sizing optimization problems.

\subsubsection{Linear Scaling}
When submitting problems to the QA, there are a number of parameters that can be fine-tuned to alter or improve the performance of the QA. The magnitude of these parameters relative to each other is typically what drives the performance of the QA. To set a consistent baseline for solving the truss sizing optimization problems, it is therefore convenient to scale the magnitude of the problem to a user-defined value. In this way, the magnitude of the truss optimization problem can be scaled with respect to other constraints. Thus, a linear scaling of each of the terms in the objective function can be performed, to ensure that the terms have a consistent and controllable magnitude. 

First, the term with the maximum absolute magnitude in the objective function, $c_{\mathit{max}}$, must be found. Then, every term in the objective function $F$ can be divided by this magnitude, to perform the linear scaling. A user-defined parameter, $c_{\mathit{user}}$, is introduced to ensure that the maximum magnitude of the terms in the objective function can be exactly specified. Thus, the linear scaling of the objective function $F$ is performed as shown in \cref{eq:LinScale}. 

\begin{equation}
F_{scaled} = \frac{c_{\mathit{user}} \times F}{c_{\mathit{max}}}
\label{eq:LinScale}
\end{equation}

Inside the linearly scaled objective function $F_{scaled}$, the term with the maximum magnitude will have a magnitude of $c_{user}$. By altering the value of $c_{\mathit{user}}$, the user is able to control the relative importance of the objective function with respect to other problem-specific parameters.

\subsubsection{Non-Linear Scaling} 
Throughout the brute-force testing of the objective functions for the different sample problems in \cref{sec:SetupObjective}, it was seen that with certain objective functions the global minimum solution and other local minimum solutions can have nearly identical function values. This was especially the case for the three-truss problem, using the fractional objective function. When the differences between solutions are small, it becomes difficult for the QA to find the global optimum solution. The main cause for very small differences between local minima and the global minimum is due to specific terms in the objective function with very small, high-precision coefficients. 

Due to analog control errors, small and high-precision coefficients are difficult for the QA to correctly take into account \cite{Dwave_TechnicalDescription}. Furthermore, when a problem is submitted to the D-Wave QA, it is by default automatically scaled (by \textit{auto\_scale}) to ensure the problem coefficients fall inside the controllable range of the QA. For linear QUBO coefficients this range is between -2 and 2. For the quadratic QUBO coefficients the maximum available range is between -2 and 1 \cite{Dwave_QPU_Properties_DW2000Q}, however, by default the QA used a range between -1 and 1. These ranges can be directly queried from the D-Wave solver by Python commands:

\makebox[\linewidth][c]{\ttfamily{DWaveSampler().properties['h\_range']}}

\makebox[\linewidth][c]{\ttfamily{DWaveSampler().properties['j\_range']}}

\noindent It is good to be mindful of this automatic scaling before submitting problems to the QA, to prevent cases where a problem is unexpectedly scaled down to a magnitude that cannot feasibly be controlled by the QA.

To assist the QA with small high-precision coefficients, it would be beneficial if these coefficients could be amplified. In turn, this may also amplify the differences between the global and other local minimum solutions. This would potentially help the QA in finding the global optimum solution. Furthermore, this may also help the QA to more effectively utilize information from the objective function, as previously insignificant terms may be amplified to a magnitude that the QA can actually take into account. To help address these issues, a non-linear scaling method was developed, with which very small coefficients in the objective function are scaled to become larger and more influential, while terms terms that are already significant are not scaled as much. 

A Python function was created and implemented that performs this non-linear scaling. The method relies on simple user-defined parameters that will allow for manual tweaking once the problem has been fully implemented for use with the QA. Given the user-defined scaling parameter $c_{\mathit{NL}}$, a positive coefficient $c_{\mathit{in+}}$ from the objective function is input into the non-linear scaling function. The non-linearly scaled coefficient $c_{\mathit{out+}}$ is then found by \cref{eq:Scaling_NL_pos}.

\begin{equation}
c_{\mathit{out+}} = \frac{c_{\mathit{in+}}}{c_{\mathit{in+}} + c_{\mathit{NL}}}
\label{eq:Scaling_NL_pos}
\end{equation}

\noindent For negative input coefficients, $c_{\mathit{in-}}$, the non-linear scaling is performed by \cref{eq:Scaling_NL_neg}, yielding the negative non-linearly scaled coefficient $c_{\mathit{out-}}$.

\begin{equation}
c_{\mathit{out-}} = -\frac{c_{\mathit{in-}}}{c_{\mathit{in-}} - c_{\mathit{NL}}}
\label{eq:Scaling_NL_neg}
\end{equation}

To determine if the input coefficient must be scaled using either \cref{eq:Scaling_NL_pos} or \cref{eq:Scaling_NL_neg}, a simple if-statement is used. By setting $c_{\mathit{NL}}$ to be a certain small number, such as 0.025, the amount of scaling that is applied to small coefficients is much more aggressive than for relatively larger coefficients. 

\begin{figure}[h!]
	\centering
		\includegraphics[width=0.80\textwidth]{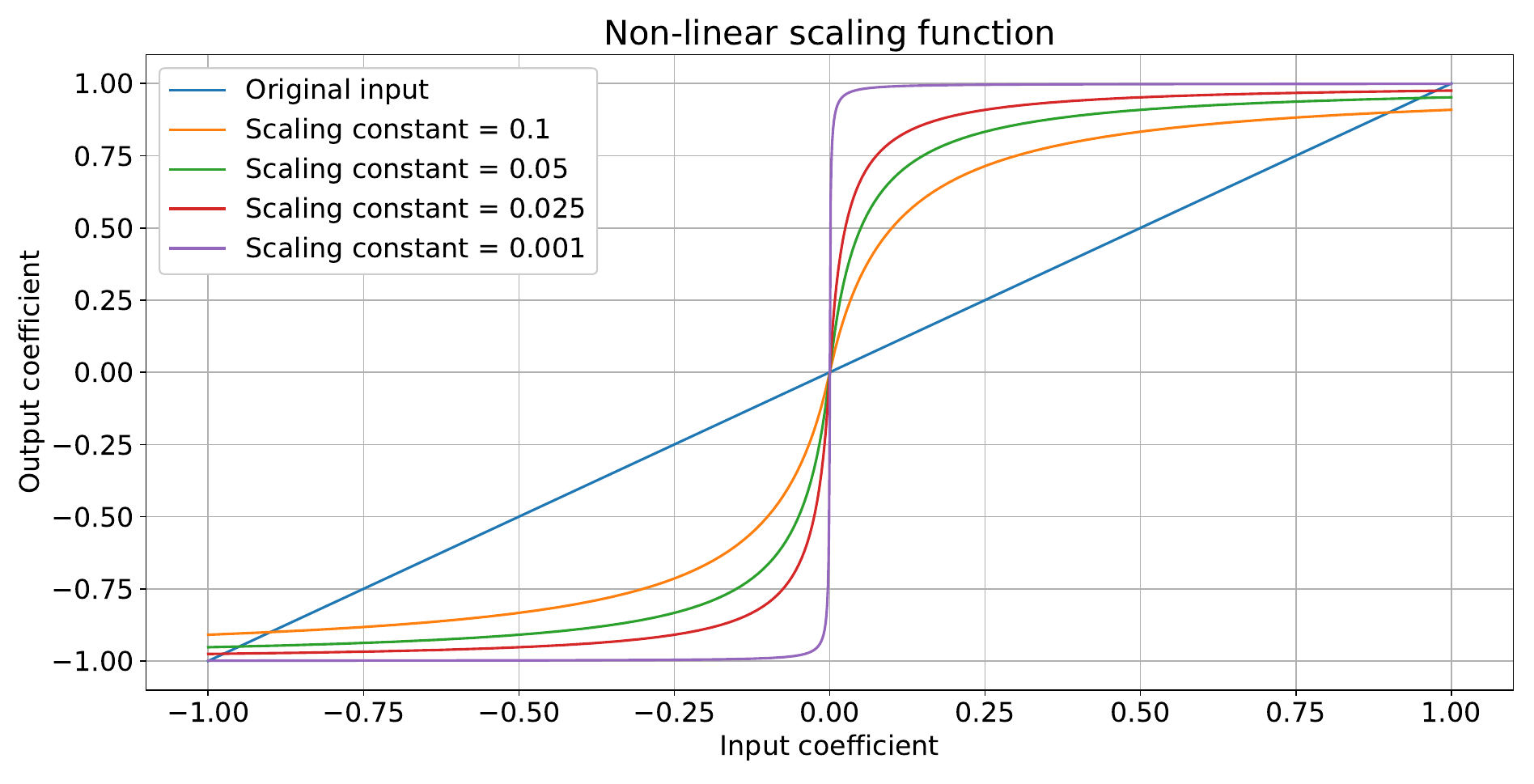}
	\caption{Non-linear scaling function for input coefficients between -1 and 1. The output of the function is shown for various values of the non-linear scaling parameter $c_{\mathit{NL}}$.}
	\label{fig:Non_linear_scaling}
\end{figure}

As a demonstration of the non-linear scaling, several plots are given in \cref{fig:Non_linear_scaling}, showing the influence of changing the parameter $c_{\mathit{NL}}$. It can be seen that for small coefficients the scaling is much more significant than for relatively larger coefficients. However, for very aggressive values of $c_{\mathit{NL}}$, such as 0.001, this can also cause all relatively large coefficients to become essentially equal. The use of this non-linear scaling function will therefore be a balancing act of increasing the importance of small coefficients, while not losing distinction for the larger terms. The Python code for the non-linear scaling function, and for producing the plot from \cref{fig:Non_linear_scaling} is available online \cite{DataverseAnalyticalQUBO}. 

The effect of the non-linear scaling is also investigated for the specific truss sizing sample problems. The non-linear scaling is performed during the iterative solving procedure. Specifically, it is performed between steps 3 and 4 of the procedure outlined in \cref{eq:Ajagekar_iter}, and is applied to the rewritten non-fractional objective function. The effect of the non-linear scaling is investigated for the first iteration of the sample problems. This first iteration is convenient to investigate, as \cref{eq:Ajagekar_iter} shows that $\lambda = 0$, meaning that the objective function in step 3 in \cref{eq:Ajagekar_iter} only involves the numerator of the original fractional objective function. 

\begin{figure}[hp]
	\centering
	\includegraphics[width=0.80\textwidth]{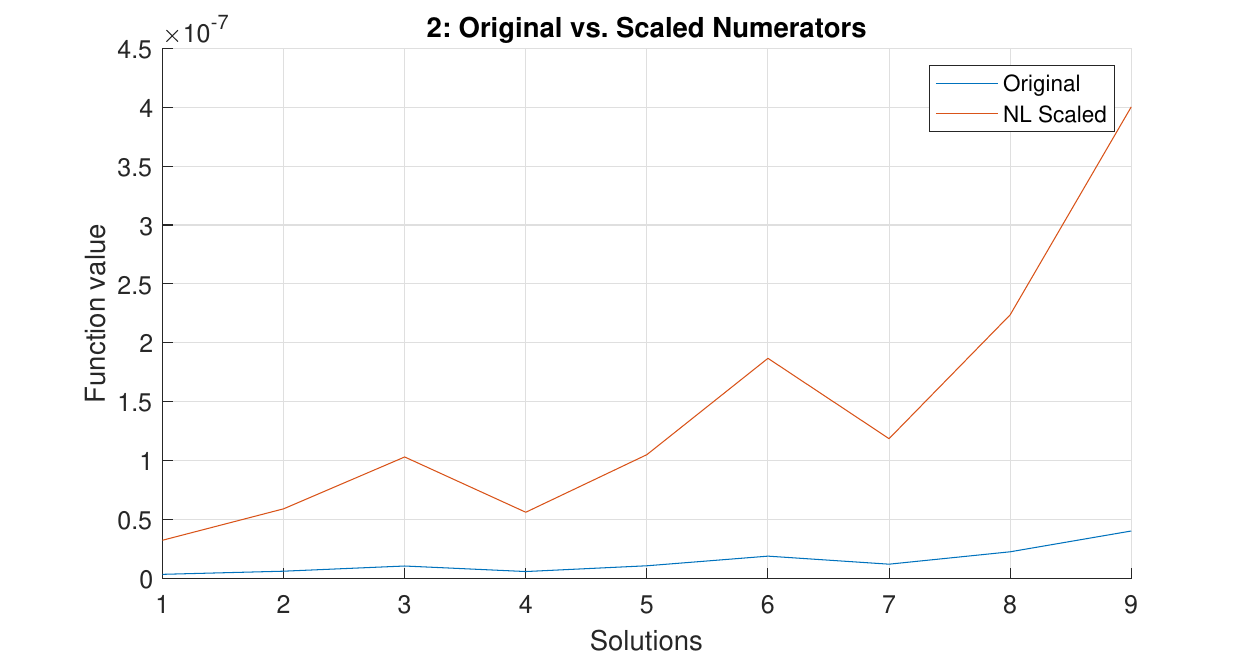}
	\caption{Effect of non-linear scaling in two-truss problem, for the first iteration of the iterative non-fractional solving method.}
	\label{fig:2Numerator}
\end{figure}

\begin{figure}[hp]
	\centering
	\includegraphics[width=0.80\textwidth]{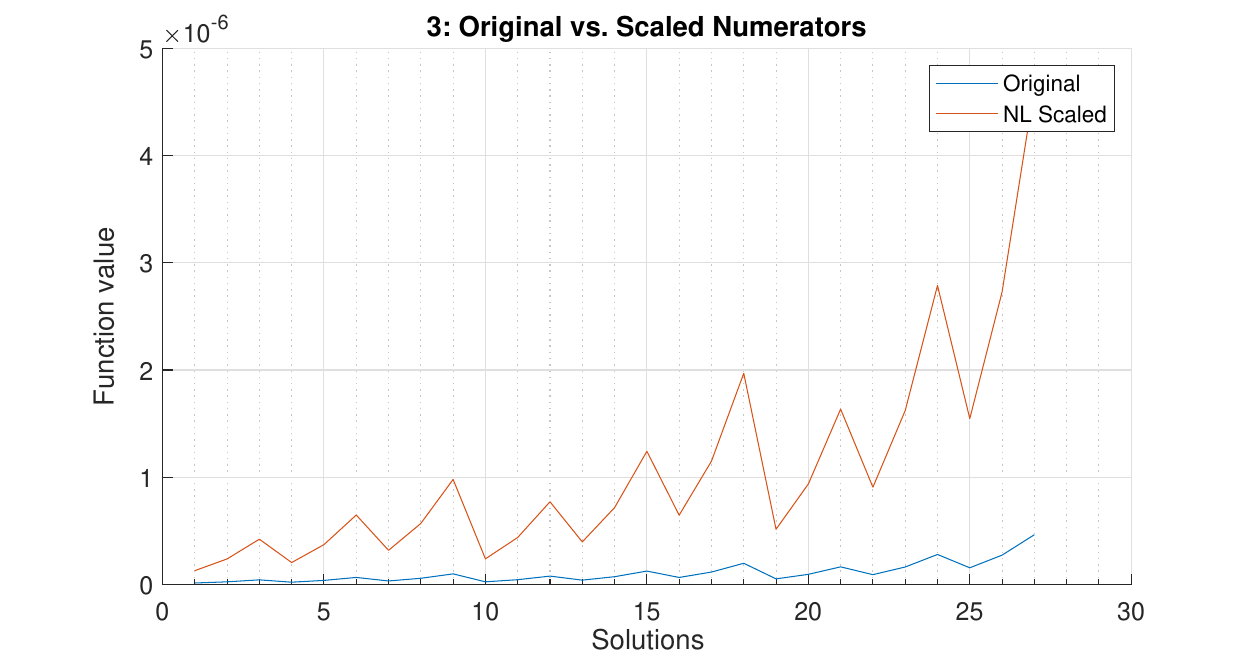}
	\caption{Effect of non-linear scaling in three-truss problem, for the first iteration of the iterative non-fractional solving method.}
	\label{fig:3Numerator}
\end{figure}

\begin{figure}[hp]
	\centering
		\includegraphics[width=0.80\textwidth]{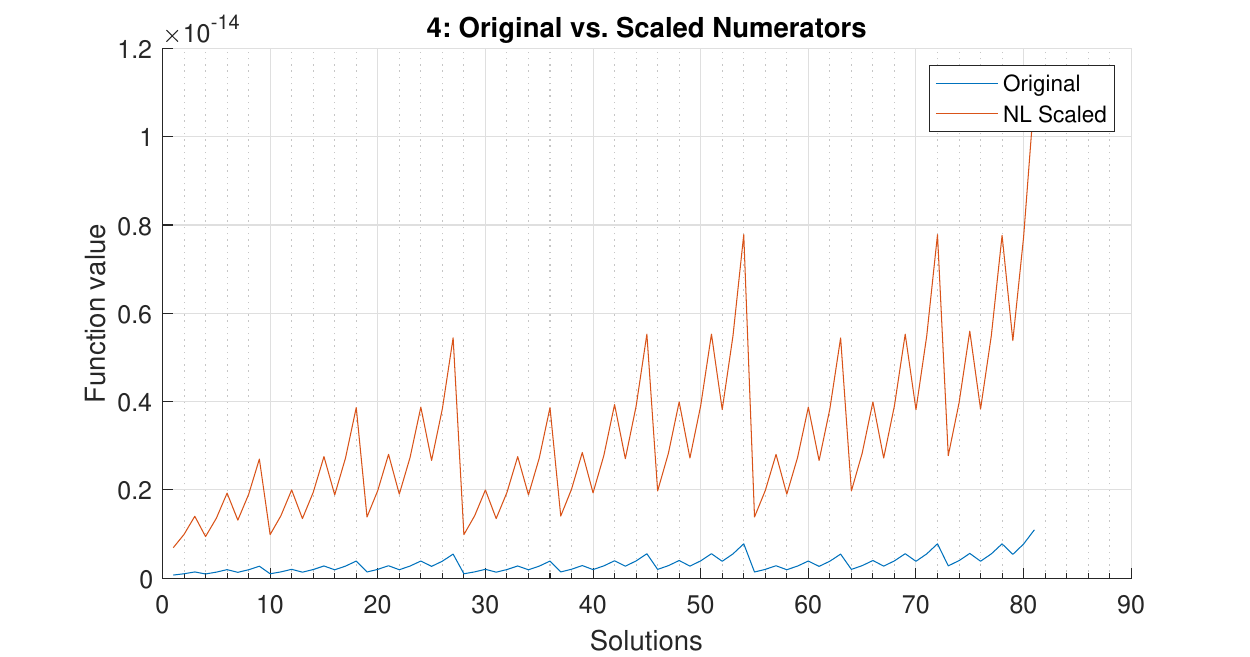}
	\caption{Effect of non-linear scaling in four-truss problem, for the first iteration of the iterative non-fractional solving method.}
	\label{fig:4Numerator}
\end{figure}

The rewritten non-fractional function is first linearly scaled, choosing the user-defined maximum coefficient to be $c_{\mathit{user}} = 1$. This brings the function values in the energy landscape to a more reasonable magnitude and allows for the non-linear scaling to work as intended. Then, the non-linear scaling can be applied, using a scaling parameter of $c_{\mathit{NL}}=0.1$. The plots in \cref{fig:2Numerator,fig:3Numerator,fig:4Numerator} show both the original (linearly scaled) and non-linearly scaled energy landscapes of the first iteration in the solving procedure for the sample truss problems. It can be seen that the small fluctuations in the energy landscape are amplified, which should make the problem easier to solve for the QA. As such, the non-linear scaling will be applied when submitting problems to the QA. 

It is important to note that this non-linear scaling function was developed solely for the purpose of increasing the differences between the global and local minima in the energy landscapes, as a part of this study. It is not intended to be used as a general-purpose tool for problems with unknown energy landscapes. Since, when using overly aggressive scaling factors for $c_{\mathit{NL}}$, this may cause the global minimum solution to change. However, this is not expected to be an issue for the truss sizing optimization problem. In step 6 of the iterative solving procedure in \cref{eq:Ajagekar_iter}, the interim solution $\hat{\textbf{q}}$ is evaluated with the original fractional objective function, meaning the iterative procedure should eventually converge on the global optimum of the original fractional objective function.

\subsubsection{Truncation of Insignificant Terms}
Once the linear and non-linear scaling has been performed it is possible to further simplify the objective function. After all scaling has been applied, certain terms in the objective function might still have a magnitude very close to zero. Such terms are difficult for the QA to take into account, as there is a finite precision with which qubit biases can be controlled within the physical hardware of the QA \cite{Dwave_TechnicalDescription}. Since the magnitude of terms in the objective function might be smaller than the precision that the QA can control, such terms cannot reliably be taken into account. This is because the analogue control-error for the qubit biases can be larger than the actual term in the objective function. Thus, terms that are too small in magnitude can simply be removed from the objective function, as trying to include them is akin to simply introducing noise into the function. A conservative truncation magnitude, beyond the precision that the QA can control, is to remove terms with magnitudes smaller than $10^{-8}$. This reduces the complexity of the objective function, which may have a positive influence on the ability of the QA to find the global optimum solution to the truss sizing optimization problem.

\subsubsection{Unary Constraint}
The majority of all possible solutions to the truss optimization problems are invalid. Solutions are invalid when an incorrect number of cross-sectional areas are selected. In other words, when either zero or more than one cross-section is selected for a truss member, the solution is invalid. Solutions can only be valid when exactly one cross-sectional area is chosen for every truss member. To promote the selection of valid solutions, the \textit{unary constraint} is implemented. 

For every truss element $n$ in a truss system, with a total of $C$ possible choices of cross-sectional area, \cref{eq:constraint_unary1} must be true for valid solutions to the truss sizing optimization problem.

\begin{equation}
\sum_{c=1}^{C}{q_{n,c}} = 1
\label{eq:constraint_unary1}
\end{equation} 

Since only one of the qubits on which this constraint acts must take a value of 1, while the others must all equal 0, this constraint is sometimes referred to as the \textit{unary constraint}. Further elaboration on the unary constraint is given in \cite{Lucas2018}. In the context of the truss optimization problem, it enforces that only one cross-section is chosen per truss element. However, in the form shown in \cref{eq:constraint_unary1}, the constraint cannot be applied in the QUBO problem framework. This is because the constraint is currently written as an equality constraint, which per definition is incompatible with quadratic \textit{unconstrained} binary optimization problems. For the constraint to become compatible with QUBO problems it must be rewritten as a minimization problem. A common method is to rewrite the equality constraint as a penalty function, using a `squared-error' approach~\cite{Kochenberger2014,Vyskocil2019}. This approach is also used, for example, in the works by \citeauthor{Vreumingen2019} and \citeauthor{Neukart2017} \cite{Neukart2017,Vreumingen2019}. Thus, the constraint can be rewritten as a minimization problem as shown in \cref{eq:constraint_unary2}.

\begin{equation}
\begin{aligned}
\left(\sum_{c=1}^{C}{q_{n,c}}\right) - 1 = 0 \\
\left(\left(\sum_{c=1}^{C}{q_{n,c}}\right) - 1\right)^{2} = 0
\end{aligned}
\label{eq:constraint_unary2}
\end{equation} 

Now, adding in a penalty scaling factor $\lambda$, the Hamiltonian energy penalty function for the unary constraint becomes:

\begin{equation}
\begin{aligned}
H_{U} = \lambda \left(\left(\sum_{c=1}^{C}{q_{n,c}}\right) - 1\right)^{2}
\end{aligned}
\label{eq:Unary_Penalty}
\end{equation} 

For the truss sizing optimization problems, only three possible choices of cross-sectional area are available per truss element, meaning that $C=3$. Therefore, \cref{eq:Unary_Penalty} can be expanded and simplified, as shown in \cref{eq:constraint_unary3}.

\begin{equation}
\begin{aligned}
H_{U} =& \lambda \left(q_{n,1} + q_{n,2} + q_{n,3} - 1 \right)\left(q_{n,1} + q_{n,2} + q_{n,3} - 1 \right)\\
H_{U} =& \lambda \left(q_{n,1}^2 + q_{n,1}q_{n,2} + q_{n,1}q_{n,3} - q_{n,1} \right. \\
& \left. + q_{n,1}q_{n,2} + q_{n,2}^2 + q_{n,2}q_{n,3} - q_{n,2} \right.\\
& \left. + q_{n,1}q_{n,3} + q_{n,2}q_{n,3} + q_{n,3}^2 - q_{n,3} \right.\\
& \left. - q_{n,1} - q_{n,2} - q_{n,3} +1 \right)
\end{aligned}
\label{eq:constraint_unary3}
\end{equation} 

Knowing that $q_{n,c} \in \left\{0,1\right\}$, which means that $q_{n,c}^2 = q_{n,c}$, the expression can be further simplified:

\begin{equation}
\begin{aligned}
H_{U} =& \lambda \left(2 q_{n,1}q_{n,2} + 2 q_{n,2}q_{n,3} + 2 q_{n,1}q_{n,3} - q_{n,1} - q_{n,2}- q_{n,3} + 1\right)
\end{aligned}
\end{equation} 

Lastly, the final constant term can be dropped, since it is independent of the qubit variables, and does not affect the minimization problem. Doing so makes the unary constraint penalty function compatible with the QUBO problem framework. Since, the penalty function can now be written as a pure summation of linear and quadratic terms, as shown in \cref{eq:Unary_sum}. 

\begin{equation}
\begin{aligned}
H_{U} =& \lambda \left(2 q_{n,1}q_{n,2} + 2 q_{n,1}q_{n,3} + 2 q_{n,2}q_{n,3} - q_{n,1} - q_{n,2}- q_{n,3} \right)
\end{aligned}
\label{eq:Unary_sum}
\end{equation} 

By adding the unary constraint, for each truss element, to the overall truss sizing objective function, the QA is more likely to find valid solutions. The strength of the unary constraint can be fine-tuned by altering the value of the user-defined parameter $\lambda$. By trail and error it has been found that a good starting value of $\lambda$ is to be twice the magnitude of the maximum term in the objective function. However, if invalid solutions are consistently found, the strength of the constraint can be increased until noncompliance with the constraint stops.

\subsubsection{Quadratization}
Up until this point, the objective function has been manipulated, scaled, and truncated in order for it to become more compatible with the QUBO problem formulation. However, one key issue has yet to be solved: the function might still contain many terms that are greater than quadratic order. Therefore, it can still not be used with the QA, since per definition the QA can only solve quadratic problems. Performing a quadratization of a high-order objective function ensures that it is rewritten as a quadratic-order function, with equivalent solutions.

There are many different methods of quadratization discussed in literature, an extensive overview of which is given by \citeauthor{Dattani2019Quadratization} \cite{Dattani2019Quadratization}. Some of these methods utilize auxiliary variables to rewrite the high-order objective function into an equivalent quadratic-order expression, while other methods are able to do so without the need for auxiliary variables. Each method has its respective benefits and drawbacks. For example, it is convenient when no auxiliary variables are needed, yet in that case it may require much effort to rewrite the objective function in an equivalent quadratic form. Alternatively, if a method uses auxiliary variables, it might be easier to find an equivalent quadratic form, but the additional variables increase the complexity of the objective function, making it more difficult to find the optimum \cite{Dattani2019Quadratization}. 

In practice, the most straightforward way to perform the quadratization is to simply rely on the implementation that is provided by D-Wave \cite{Dwave_make_quadratic}. In their implementation, all high-order terms are rewritten and replaced to be in terms of auxiliary variables, such that the final problem is at most of quadratic order. An additional user-defined parameter is used to select the strength with which the quadratization is enforced \cite{Dwave_make_quadratic}. If the quadratization is not enforced correctly, this can result in a poor approximation of the original high-order objective function. The quadratization strength is problem-dependent and must be tuned such that the quadratization is always obeyed, to have an accurate representation of the original high-order objective function.

With the quadratization, the discrete truss sizing optimization problem is finally written as a QUBO problem. This vital step finally concludes all of the preparatory work that was necessary for the problem to be compatible with the quantum computing hardware. The next section will discuss the solution process for the discrete truss sizing optimization problems.

\subsection{Solving the Discrete Truss Sizing QUBO}
Having completed the previous processing steps, the truss sizing optimization problems can now finally be written in a QUBO formulation and can be solved using the QA. In this section, some notable parameters that influence the behavior of the QA are discussed, as well as the testing approach for understanding the performance of the QA for the three sample problems.

\subsubsection{Overview of Analyses}
To solve the reference truss problems, two different analysis methods are applied. First, brute-force evaluation of the original fractional objective function is used to obtain a baseline solution. Then, quantum annealing is used, following the procedures described in the previous sections. The specifications of the local classical computing hardware that was used in this research are given in \cref{tab:computer}.

\begin{table}[htbp]
\caption{Classical computing hardware.}
\begin{tabular}{|l|l|} \hline
Device & Lenovo Legion Y540-15IRH      \\ \hline
CPU    & Intel Core i7-9750H 2.6 GHz \\ \hline
Memory & 16 GB DDR4 2667 MHz           \\ \hline
GPU    & Mobile NVIDIA RTX 2060 6GB   \\ \hline
\end{tabular}
\label{tab:computer}
\end{table}

The brute-force analyses will simply serve to obtain a reference solution to the truss sizing optimization problems. There are other more efficient (and more complicated) classical analysis methods available for truss optimization problems, as reviewed by \citeauthor{Stolpe2016} \cite{Stolpe2016}. However, for the purposes of this study, simple brute-force analysis is sufficient to produce the reference solutions. Each of the brute-force analyses will be performed three times, so that an average solve time can be calculated.

The main focus for the QA analyses will be to find the computational time and the probability of obtaining the global optimum solution. It will also be measured how long the symbolic finite element approach takes to set up the original fractional objective function for the truss sizing optimization problems. The goal will be to find out whether it is feasible to apply quantum computing to these practical truss optimization problems. Furthermore, these analyses will show whether the methods outlined in the previous sections provide a feasible means of translating the truss sizing problem to a QUBO format. To calculate the average computational time and show a basic probability for finding the global optimum solution, the QA analyses are each performed ten times. The number of analyses could unfortunately not be increased due to limitations placed on the amount of quantum computational time allotted to basic user accounts on the D-Wave Leap platform.

\subsubsection{Parameter Tuning}
In the previous sections, a number of user-defined parameters were introduced that influence the objective functions of the three truss sizing problems. Values need to be chosen for these parameters before the problems can be submitted to the QA. Overall, the following parameters all need to be given values:

\begin{itemize}
    \item Iterative solving procedure
    \begin{itemize}
        \item Maximum number of iterations allowed
        \item Iteration convergence threshold
    \end{itemize}
    
    \item Objective function processing
    \begin{itemize}
        \item Highest order terms allowed
        \item Linear scaling magnitude
        \item Non-linear scaling strength
        \item Precision truncation magnitude
        \item Unary contraint strength
        \item Quadratization strength
    \end{itemize}

    \item Quantum annealing
    \begin{itemize}
        \item Number of reads
        \item Chain strength
    \end{itemize}
\end{itemize}

To aid in finding sensible values for most of the above parameters, without wasting the limited amount of quantum computational time available, an alternative classical solver is used. D-Wave provides a Simulated Annealing (SA) solving algorithm, which, similar to their quantum annealer, can be used to solve QUBO problems. The main difference is that the SA solver only relies on the local classical computing hardware, and does not expend any of the quantum computational time allowance. The SA solver was therefore used to test the functionality of the Python code, and to find initial values for the relevant problem parameters.  

Values for the parameters related to the iterative solving procedure were first chosen. Testing via initial trial SA analyses showed that around 5 iterations are needed to find a converged solution. This value will be tripled for the quantum annealing analyses to give some room for potential errors to be corrected during the procedure. Thus, the maximum number of iterations within one solving attempt will be set to 15. If the procedure does not converge after the maximum number of iterations is reached, the analysis is stopped, to prevent excessive expending of computational time. The convergence threshold $\delta$ for the iterative procedure (as seen in \cref{eq:Ajagekar_iter}) is set at a value of $10^{-6}$, which is the same value used by \citeauthor{Ajagekar2019} 

Starting values for the different objective function processing parameters were also found from initial trial SA analyses. First of all, the highest order of terms allowed in the objective function for each of the sample problems will be set equal to the number of truss elements. Second, it was chosen to set the linear scaling maximum magnitude $c_{\mathit{user}}$ to a value of 1 for all analyses. Third, the non-linear scaling parameter $c_{\mathit{NL}}$ has been set to a value of 0.1 for all analyses. During brute-force testing it was seen that this value improves the distinction between minima in the energy landscapes for the sample problems, while preserving the global minima. Fourth, the unary constraint strength was set to a value of 10 for the two- and three-truss problems. For the four-truss problem it was set to a value of 20. Testing via SA showed that with these settings the constraint was obeyed consistently, yielding valid solutions to the truss sizing problems. Fifth, terms that have a magnitude smaller than $10^{-8}$ are truncated, to slightly reduce the number of terms in the objective functions. This is a conservative truncation, as it is well beyond the precision that the QA hardware can control. Finally, the quadratization strength is set to a value of 10 for the two- and three-truss problems, and a value of 20 for the four-truss problem, such that it equally matches the strength of the unary constraint.

For the QA, some additional parameters are needed to solve the sample problems. The \textit{number of reads} describes the number of times a specific problem is solved by the QA, before the final best-performing solution is returned to the user. Increasing the number of reads increases the likelihood of obtaining optimal solutions, at the expense of additional computational time. To minimize the expenditure of quantum computational time, a viable number of reads was first estimated using the SA solver. 

Using the SA solver, some tests were performed in which the number of reads was set to values of 16, 64, and 256, for each of the three sample problems. For the two-truss problem, the global optimum results were obtained every single time, regardless of the number of reads. For the three-truss problem, using 64 and 256 reads reliably gave optimal or near-optimal results. For the four-truss problem, only when setting the number of reads to 256 was the global optimum result reliably obtained. 

The settings for the number of reads for the QA analyses were chosen based on the testing that was done using SA. Namely, for the two-truss problem, the number of reads will be set to 16 and 64. For the three-truss problem the number of reads will be set to 64 and 256. Finally, for the four-truss problem, the number of reads will only be set to 256. 

The \textit{chain strength} parameter for the QA relates to a physical issue called \textit{chain break}, which can occur while the QA is solving problems. To explain the fine-tuning of the chain strength parameter, some additional context will first be given to explain the chain break phenomenon. 

A QUBO problem is defined through an upper-triangular $N \times N$ matrix $\textbf{Q}$. The terms on the diagonal of the matrix $\textbf{Q}$ relate only to a single problem variable, while the off-diagonal terms describe an interaction between two different problem variables. On the D-Wave 2000Q, which was used during this research, each qubit can only directly interact with at most six other qubits \cite{Dwave_gettingstarted}. This limited connectivity between qubits can present an issue for larger QUBO problems. 

When a QUBO is submitted to the QA, it must be \textit{embedded} onto the physical architecture of the QPU. The embedding translates the logical structure of a QUBO problem and translates it to the physical structure of the QA. This also means that the logical problem variables are translated to physical qubits on the QPU. However, when the QUBO problem necessitates a high connectivity between logical problem variables it can become impossible to directly embed every logical variable onto individual single qubits. By \textit{chaining} together strings of multiple qubits, and enforcing them to act as single logical problem variables, the connectivity between the embedded logical variables can be increased beyond the limitations of the physical hardware \cite{Fang2019}. 

In practice, problems can become very challenging for the QA to solve if the number of interactions between variables is high. A QUBO problem with high connectivity requirements will lead to embeddings with very long qubit chains. In turn, the longer the qubit chains are, the more difficult it is to enforce the chained qubits to act in unison. When a qubit chain is working as intended, the chained qubits will all end up in the same final ground state, all being either 0 or 1. However, when a qubit chain is \textit{broken} the qubits in the chain end up with a mix of 0 and 1 states. This makes the final state of the qubit chain, and the intended final state of the corresponding logical problem variable unclear. By increasing the chain strength, the risk of chain breaks occurring is reduced. However, overly high chain strength should be avoided as the relative strength of the objective function itself is reduced, potentially making it more difficult to obtain the true global optimum solution. 

Initially, a chain strength of 10 was selected, but it was seen that chain breaks would still occur for the three- and four-truss problems. For those problems a chain strength of 30 performed better, preventing chain breaks from occurring. There are very many additional parameters that can be tuned to alter the performance of the quantum annealer, but these were left in their default configurations \cite{Dwave_solver_properties}. 

In \cref{tab:analysis_parameters} a summary is given of all parameters used for each of the analyses. Parameters that are irrelevant or not applicable for an analysis are indicated by \textit{NA}. 

\begin{table}[htbp]
\caption{Parameters used for all analyses.}
\label{tab:analysis_parameters}
\makebox[\linewidth]{
\footnotesize
\renewcommand{\arraystretch}{1.35} 
\begin{tabular}{|l|ccc|ccc|ccc|}
\hline
Parameters & \multicolumn{3}{c|}{Brute-force} & \multicolumn{3}{c|}{Quantum Annealing} \\ \hline
Truss system & 2-truss & 3-truss & 4-truss & 2-truss & 3-truss & 4-truss\\ \hline
Total number of times analyzed & 3 & 3 & 3 & 10 & 10 & 10               \\ \hline
Maximum number of iterations & NA & NA & NA & 15 & 15 & 15 \\ \hline
Iteration convergence threshold & NA & NA & NA  & $10^{-6}$ & $10^{-6}$ & $10^{-6}$ \\ \hline
Number of reads per iteration  & NA  & NA & NA & $\left\{16, 64 \right\}$ & $\left\{64, 256 \right\}$ & $\left\{256 \right\}$ \\ \hline
Highest order terms allowed & NA & NA & NA & 2 & 3 & 4   \\ \hline
Linear scaling maximum magnitude & NA & NA & NA & 1 & 1 & 1 \\ \hline
Non-linear scaling strength & NA & NA & NA & 0.1 & 0.1 & 0.1 \\ \hline
Unary constraint strength & NA & NA & NA & 10 & 10 & 20 \\ \hline
Quadratization strength & NA & NA & NA & 10 & 10 & 20 \\ \hline
Precision truncation magnitude & NA & NA & NA & $10^{-8}$ & $10^{-8}$ & $10^{-8}$ \\ \hline
Chain strength & NA & NA & NA & 10 & 30 & 30 \\ \hline
Annealing time $\left[\mu s\right]$ & NA & NA & NA & 20 & 20 & 20 \\ \hline
\end{tabular}}
\end{table}

\section{Results}

\subsection{Results: Two-Truss Problem}
The two-truss problem is the simplest of the three sample problems. To set up the original fractional objective function for the problem, the symbolic finite element approach is used. For this problem, it took approximately 13.5 seconds to set up the fractional objective function. Once this objective function is found, it is written to a text file. The objective function is then imported and interpreted before performing every analysis, and no longer needs to be set up from scratch. This saves some time when performing the three brute-force analyses, and for the QA, where ten analyses are performed for each different setting of the number of reads. 


For the brute-force analysis, evaluating the fractional objective function for valid solutions, an average of 0.234 seconds was needed to find the global optimum solution. The baseline global optimum solution is $[0, 0, 1, 1, 0, 0]$. This indicates the largest possible cross-sectional area should be chosen for the first truss element, while for the second truss element the smallest cross-sectional area should be chosen.

\begin{figure}[htbp]
	\centering
	\makebox[\linewidth][c]{
		\includegraphics[width=1.2\textwidth]{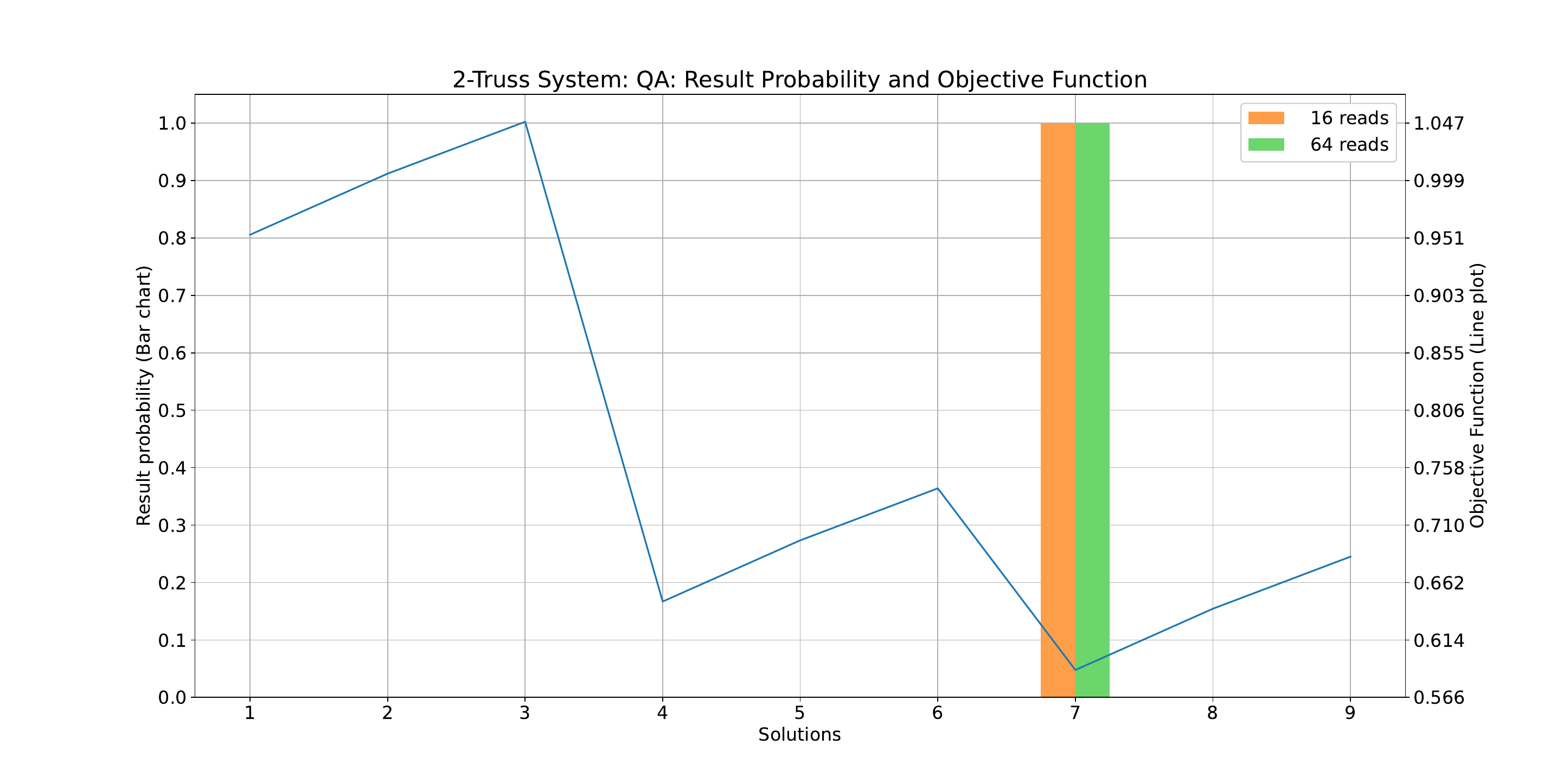}}
\caption{Solution probability histogram of two-truss problem using the QA, compared to original objective function. Global optimum solution located at solution number 7.}
	\label{fig:2TrussResultsQA}
\end{figure}

Using the QA, the analyses were performed ten times, to gain some insight into the probability of obtaining the globally optimal solution. In \cref{fig:2TrussResultsQA}, the solutions which were found by the QA are presented in a histogram. Superimposed on the histogram is a line-plot of the original fractional objective function, which had previously been calculated by brute force. 

It can be seen from \cref{fig:2TrussResultsQA} that, regardless of the choice for the number of reads, the QA finds solution number 7, which corresponds to the same global optimum solution as is obtained via the brute-force analyses.

When it comes to the computational time for the QA analyses, the following results were obtained:

\begin{itemize}
    \item With number of reads = 16
    \begin{itemize}
        \item Average total time = 19.52 s. Standard deviation = 5.10 s. 
        \item Average QPU time = 59176 $\mu$s. Standard deviation = 15061 $\mu$s. 
    \end{itemize}
    
    \item With number of reads = 64
    \begin{itemize}
        \item Average total time = 19.06 s. Standard deviation = 0.21 s.
        \item Average QPU time = 118207 $\mu$s. Standard deviation = 10.7 $\mu$s.
    \end{itemize}
\end{itemize}

\noindent From these results it is seen that setting the number of reads to 64 leads to more consistent analysis times, compared to using only 16 reads. On the other hand, by setting the number of reads to 64, the amount of QPU access time is increased significantly.

\subsection{Results: Three-Truss Problem}
The three-truss problem is a more complicated problem to set up compared to the two-truss problem, due to the increased number of variables. Setting up the fractional objective function for this problem, and writing that expression to a text file took approximately 81.4 seconds. Once the expression was written to a file, it could simply be imported and reused for every analysis. Overall, three brute-force analyses were performed for the three-truss problem, as well as ten QA analyses for each setting of the number of reads. 


Considering the brute-force analysis of the fractional objective function, it took an average of 0.9 seconds to find the global optimum solution $[0, 0, 1, 1, 0, 0, 0, 0,1]$. This solution indicates that the largest cross-sectional area should be chosen for the first and third truss elements, and that the smallest cross section should be chosen for the second truss element. 

\begin{figure}[htbp]
	\centering
	\makebox[\linewidth][c]{
		\includegraphics[width=1.2\textwidth]{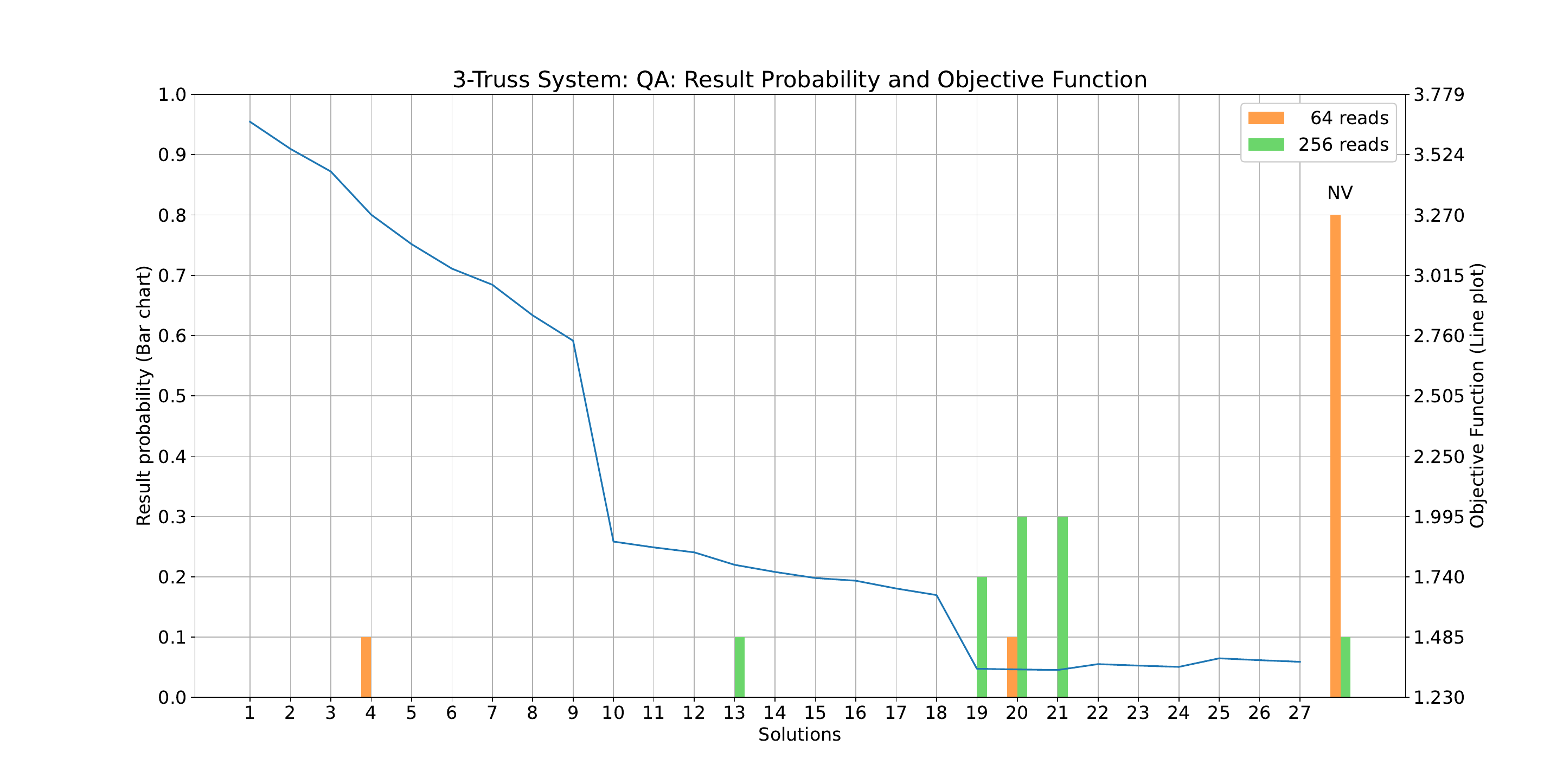}}
	\caption{Solution probability histogram of three-truss problem using the QA, compared to original objective function. Global optimum solution located at solution number 21.}
	\label{fig:3TrussResultsQA}
\end{figure}

With the QA, the analyses were performed with the number of reads set to values of 64 and 256. The solution probability histogram in \cref{fig:3TrussResultsQA} shows the results that were obtained. The figure also shows the original fractional objective function, as was calculated via brute-force. Analyses that ended with non-valid results are indicated by the 'NV' column in the histogram. 

From \cref{fig:3TrussResultsQA}, it can be seen that there is a large difference in performance, depending on the setting for the number of reads. Setting the number of reads to 64 leads to non-valid solutions for most analyses. At this setting, the QA does not reliably produce usable results. However, when the number of reads is increased to 256, optimal or near-optimal valid solutions are obtained most of the time. From these results it is also clear that the problem is more difficult for the QA to solve compared to the two-truss problem. 

As part of the QA analyses, the following computational times were measured:
\begin{itemize}
    \item With number of reads = 64
    \begin{itemize}
        \item Average total time = 98.7 s. Standard deviation = 37.5 s. 
        \item Average QPU time = 337477 $\mu$s. Standard deviation = 117808 $\mu$s. 
    \end{itemize}
    
    \item With number of reads = 256
    \begin{itemize}
        \item Average total time = 48.8 s. Standard deviation = 20.6 s.
        \item Average QPU time = 558414 $\mu$s. Standard deviation = 244331 $\mu$s.
    \end{itemize}
\end{itemize}

It is seen that setting the number of reads to 256 leads to more consistent behavior, as evidenced by the smaller standard deviation for the total analysis time. Furthermore, the higher number of reads allows for the analysis to converge more quickly, as fewer iterations are generally needed to find a solution. However, the consequence of choosing a higher number of reads is that the amount of QPU access time is also increased.

\subsection{Results: Four-Truss Problem}
The last problem, with the highest number of variables involved, is the four-truss problem. Using the symbolic finite element method, it took approximately 3431 seconds to set up the fractional objective function. This is a very significant increase compared to the roughly 81 seconds that were needed to set up the objective function of the three-truss problem. In this case, it is a great benefit that the fractional objective function is written to a text file. Testing of the analysis procedures is much faster when the objective function text file can simply be imported and interpreted, compared to prefacing every analysis by a nearly hour-long setup process. Overall, the brute-force analysis was performed three times, and the QA analysis was performed ten times with the number of reads set to 256. 


Via brute-force analysis, the global optimum solution $[1, 0, 0, 1, 0, 0, 0, 0, 1, 1, 0, 0]$, was found in an average of 4.4 seconds. This solution indicates that the first, second, and fourth truss elements should optimally use the smallest choice for the cross-sectional area. The third truss element should use the largest available cross-sectional area. 

\begin{figure}[htbp]
	\centering
	\makebox[\linewidth][c]{
		\includegraphics[width=1.2\textwidth]{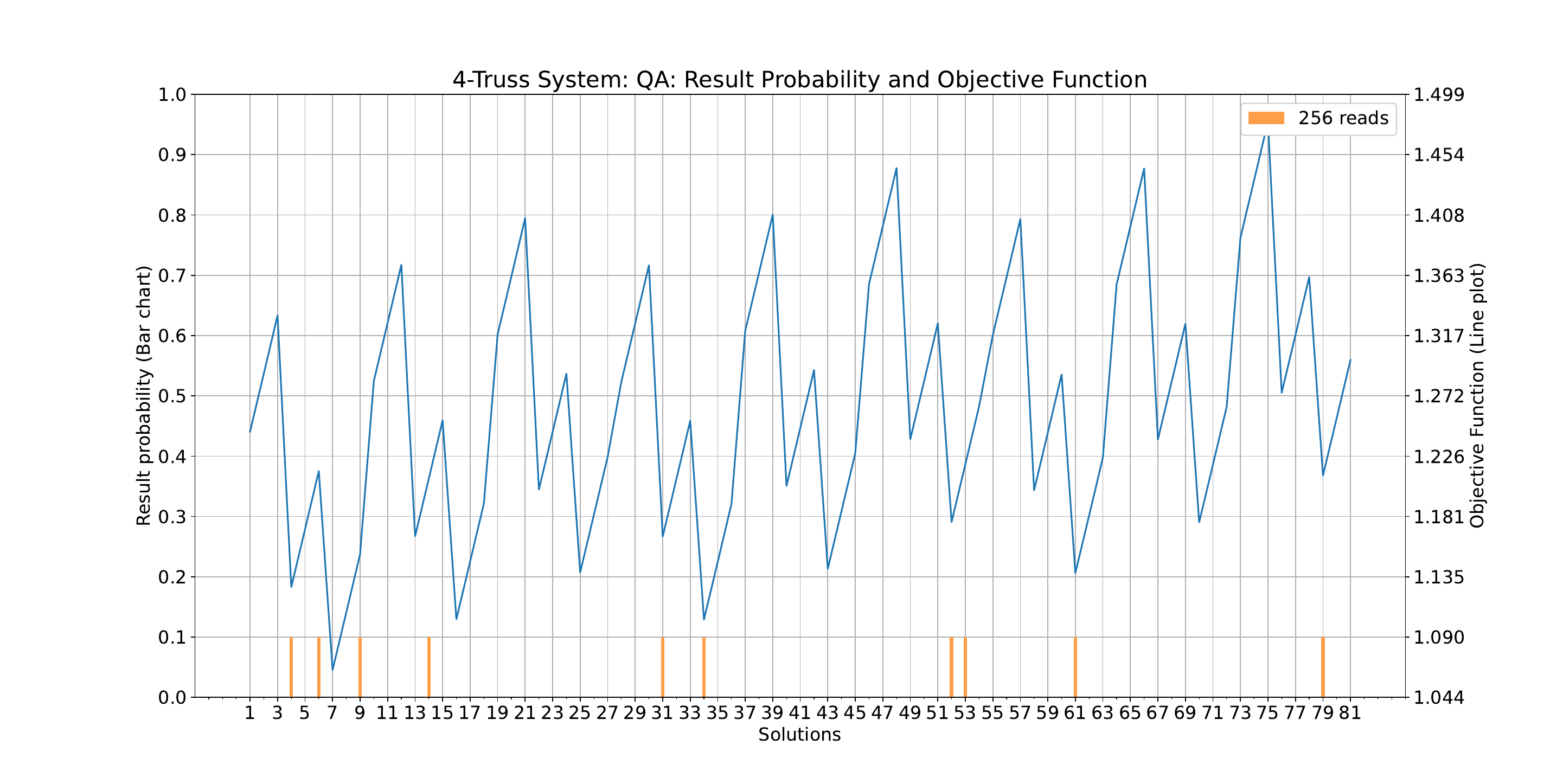}}
	\caption{Solution probability histogram of four-truss problem using the QA, compared to original objective function. Global optimum solution located at solution number 7.}
	\label{fig:4TrussResultsQA}
\end{figure}

The histogram of the results that were obtained by the QA are shown in \cref{fig:4TrussResultsQA}. The figure also shows a line plot of the original fractional objective function, which was determined by the brute-force analyses. 

From \cref{fig:4TrussResultsQA}, it is seen that seemingly random, but valid, solutions are obtained by the QA. Only one solution was found that comes within 5\% of the global optimum solution. Namely, solution number 34 was found once and has an an objective function value of 1.103. The global optimum solution is located at solution number 7, and has an objective function value of 1.065. It is worth noting that seven out of the ten analyses ended with an iteration run-out, rather than ending with a converged final solution. 

The computational time for the QA analyses was obtained: 

\begin{itemize}
    \item With number of reads = 256
    \begin{itemize}
        \item Average total time = 437 s. Standard deviation = 39.0 s. 
        \item Average QPU time = 1280156 $\mu$s. Standard deviation = 251109 $\mu$s. 
    \end{itemize}
\end{itemize}

The average time to solve is approximately nine times longer than that for the three-truss problem, using the same number of reads. Furthermore, the average of 1.28 seconds of QPU access time per analysis meant that any further testing was not possible for this problem, as this would consume too much of the 60-second monthly QPU access budget.

\section{Conclusions}
This work has established a new method to apply QA to the discrete optimization of truss structures. A symbolic FEM approach is employed to express the objective function in terms of qubit variables. As the qubit variables encode the design choices of the truss structure, they would inevitably take part in the formulation of the stiffness matrix of the structure, which leads to a fractional form for the objective function. \citeauthor{Ajagekar2019}'s iterative approach is used to approximate the original fractional objective function with a non-fractional one such that it can be made compatible with QA. The proposed method has been applied on three different discrete truss sizing optimization problems. It is found that the QA is able to find the global minimum solution if sufficient reads can be afforded. However, there are several challenges that need to be addressed for this approach to be scalable to larger problems. 

\subsection{Symbolic Finite Element Method}
The first step to translating the truss sizing problems to a compatible format is to define an objective function, written in terms of binary variables. For this purpose the symbolic finite element method was used, which was implemented in Matlab. The time to setup the objective function for each of the three sample problems was measured, and is reiterated in \cref{tab:setup_time}.

\begin{table}[htbp]
\centering
\caption{Overview of objective function setup time.}
\label{tab:setup_time}
\begin{tabular}{|c|c|c|c|}\hline
Truss System & Variables & Setup Time {[}s{]} & Growth Factor \\ \hline
2            & 6         & 13.504             & -                     \\ \hline
3            & 9         & 81.390             & 6.0270                \\ \hline
4            & 12        & 3430.542           & 42.1494               \\ \hline
\end{tabular}
\end{table}

The timings given in \cref{tab:setup_time} show that the symbolic finite element method implementation scales poorly as the number of problem variables increases. Even if the classical computing hardware used for this method improves, this method is likely infeasible for larger problems. While the symbolic finite element method allows the practical application of quantum annealing to be investigated for simple truss sizing problems, the current implementation cannot be directly applied to larger and more realistic truss structures. Further research is needed to investigate more efficient ways of solving symbolic system equations. 

\subsection{Fractional Objective Function}
When the symbolic finite element method is used to set up an objective function for a discrete truss sizing optimization problem, the resulting objective function has a fractional form. This further complicates the journey to achieving a QUBO compatible form. An iterative method was implemented that rewrites the original fractional objective function into a non-fractional form. This non-fractional function can then be further processed in order to achieve the final QUBO problem. However, a number of points make the current implementation difficult to use on the QA hardware. 

Firstly, the objective function that is obtained from the symbolic finite element method contains all possible multiplications between the available binary variables. This means that the problem initially requires an all-to-all connectivity between qubits, if it was to be directly embedded on the QPU. This research describes a number of steps that were taken to reduce the complexity of the truss sizing problem, such as removing excessively high-order terms, and truncating insignificant terms. Nevertheless, the connectivity requirements for the truss sizing problems remain far beyond the natural capability of the QA, meaning that long qubit chains are needed to embed the truss sizing problems. Not only do these long qubit chains have a negative impact on the performance of the QA, but larger truss sizing problems could eventually become infeasible to embed on the QPU, as the problem requirements could easily outgrow the physical capabilities of the QA hardware.

Secondly, as the symbolic finite element method yields a fractional objective function, the current approach relies on an iterative scheme in order to translate the objective function to a QUBO format. This means that various sources of overhead are repeatedly added to the solving process: repeatedly processing new non-fractional objective functions, waiting for an embedding to be calculated, awaiting your turn in the D-Wave problem submission queue, waiting to obtain the results from D-Wave, and repeatedly going through potentially inefficient Python programming. All of this contributes to the long total solve time, particularly for larger problems. 

\subsection{Quantum Annealing}
When finally the objective function is successfully translated to a QUBO compatible form, it is submitted to the D-Wave QA to obtain a solution. From the results obtained for the three sample problems, it is seen that the QA has increasing difficulty in finding optimal or near-optimal solutions as the problem gets larger. For the largest sample problem, the optimum solution was not found within the allowable QPU time. More reads and longer QPU calculation time will be expected to find the solution.\\

Overall, the method proposed in this research constitutes a proof-of-concept in using QA for discrete structural optimization. The concepts of setting up a symbolic objective function, finding ways to simplify an objective function, and eventually translating a problem to a QUBO format may also be applicable to other optimization problems on QA, particularly when the qubit variables form part of the stiffness matrix and a fractional objective function needs to be evaluated. The identified challenges requires further research effort to improve its scalability onto larger problems.





\vspace{6pt} 



\authorcontributions{Conceptualization, K. Wils and B. Chen; methodology, K. Wils and B. Chen; software, K. Wils; validation, K. Wils and B. Chen; formal analysis, K. Wils; investigation, K. Wils; resources, K. Wils; data curation, K. Wils; writing---original draft preparation, K. Wils; writing---review and editing, B. Chen; visualization, K. Wils; supervision, B. Chen; project administration, B. Chen; All authors have read and agreed to the published version of the manuscript.}

\funding{This research received no external funding.}

\institutionalreview{Not applicable.}

\informedconsent{Not applicable.}

\dataavailability{All Python and Matlab code produced during this research, as well as the results that were obtained, are publicly available \cite{DataverseAnalyticalQUBO}.} 

\acknowledgments{The authors would like to acknowledge the intellectual discussions with Mr. Giorgio Tosti Balducci, PhD candidate in the same department, during this project.}

\conflictsofinterest{The authors declare no conflict of interest.}



\abbreviations{Abbreviations}{
The following abbreviations are used in this manuscript:\\

\noindent 
\begin{tabular}{@{}ll}
MDPI & Multidisciplinary Digital Publishing Institute\\
DOAJ & Directory of open access journals\\

QA & Quantum Annealer\\
GPQC & General Purpose Quantum Computer\\
QPU & Quantum Processing Unit\\
QUBO & Quadratic Unconstrained Binary Optimization\\
FEM & Finite Element Method\\
SA & Simulated Annealing\\
\end{tabular}
}





\begin{adjustwidth}{-\extralength}{0cm}

\reftitle{References}


\bibliography{bibliography}

%


\PublishersNote{}
\end{adjustwidth}
\end{document}